\documentclass[12pt]{article}
\usepackage[latin9]{inputenc}
\usepackage{color}
\usepackage{amsmath}
\usepackage{amssymb}
\usepackage{graphicx}

\makeatletter

\newcommand{\lyxmathsym}[1]{\ifmmode\begingroup\def\b@ld{bold}
  \text{\ifx\math@version\b@ld\bfseries\fi#1}\endgroup\else#1\fi}

\providecommand{\tabularnewline}{\\}

\usepackage{amsfonts}

\usepackage{epsfig}

\makeatother

\begin{document}
\begin{titlepage} 

\begin{center}
\textbf{\LARGE The 4-Body Problem in a (1+1)-Dimensional Self-Gravitating
System}\textbf{ }\\
\textbf{$ $}
\par\end{center}

$ $

\begin{center}
\textbf{A. Lauritzen\footnotemark\footnotetext{email: andrew.t.lauritzen@intel.com},
P. Gustainis\footnotemark\footnotetext{email: pgustain@uwaterloo.ca},
and R.B. Mann\footnotemark\footnotetext{email: mann@avatar.uwaterloo.ca}
}\\
\textbf{ \vspace{0.5cm}
 Dept. of Physics \& Astronomy, University of Waterloo Waterloo, ONT
N2L 3G1, Canada}\\
\textbf{ \vspace{2cm}
 }\\
\textbf{ \vspace{2cm}
 \today}\\
\textbf{ }
\par\end{center}
\begin{abstract}
We report on the results of a study of the motion of a four particle
non-relativistic one-dimensional self-gravitating system. We show
that the system can be visualized in terms of a single particle moving
within a potential whose equipotential surfaces are shaped like a
box of pyramid-shaped sides. As such this is the largest $N$-body
system that can be visualized in this way. We describe how to classify
possible states of motion in terms of Braid Group operators, generalizing
this to $N$ bodies. We find that the structure of the phase\textcolor{black}{{}
space of each of these systems yields a large variety of interesting
dynamics, containing regions of quasiperiodicity and chaos. Lyapunov
exponents are calculated for many trajectories to measure stochasticity
and previously unseen phenomena in the Lyapunov graphs are observed.}
\end{abstract}
\end{titlepage}\newpage{}

\section{Introduction}

\label{sec:intro} One of the oldest problems in physics is that of
determining the motion of $N$ bodies under a specified mutual force.
Commonly referred to as the $N$-body problem, it occurs frequently
in many distinct subfields and remains an active area of research.
\ When the specified interaction is gravitation the problem is particularly
interesting, partly because of obvious astrophysical applications
and partly because some basic issues in the statistical behaviour
of such systems are still not well-understood.

One-dimensional self-Gravitating Systems (OGS's) continue to play
an important role in this regard. \ Even in the simplified setting
of one spatial dimension, there are still many open questions about
the OGS concerning its ergodic behaviour, the conditions (if any)
under which equipartition of energy is attained, and whether or not
it can reach a true equilibrium configuration from arbitrary initial
conditions. Furthermore, even for non-relativistic (Newtonian) gravity,
OGS's have proven to be very useful in modeling many diverse physical
systems. Stable core-halo structures have been shown to exist in the
OGS phase-space that are reminiscent of those found in globular clusters
\cite{yawn}, in which a dense core of particles near equilibrium
are surrounded by a cloud of particles with high kinetic energy that
interact very weakly with the core. The OGS also models the motion
of stars interacting with a highly flattened galaxy \cite{Rybicki}\ and
the dynamics of flat, parallel sheets colliding along a perpendicular
axis \cite{LMiller}. A preliminary study of the relativistic case
yielded a complete derivation of the partition and single-particle
distribution functions in both the canonical and microcanonical ensembles
to leading order in a post-Newtonian expansion \cite{pchak}. \ Recently
non-relativistic OGS's have been shown to exhibit a new phase of evolution
in which fractal spatial structure emerges from non-fractal initial
conditions \cite{fractal}.

Even for small values of $N$, OGS's exhibit interesting novel behaviour
and model interesting physical systems. The 3-body OGS is equivalent
to a system of two elastically colliding billiard balls in a uniform,
gravitational field \cite{Goodings}, as well as to a bound state
of three quarks to form a ``linear baryon'' \cite{Bukta}. \ It
can be extended to fully include relativistic gravitational interactions
\cite{fionashort}, and investigations have been carried out for both
the equal mass \cite{Burnell} and unequal mass \cite{justinrobb}
cases . \ Furthermore, it is isomorphic to a system in which a billiard
elastically collides with a wedge of fixed angle in a uniform, gravitational
field \cite{LMiller}. \ As such, one can study the 3-body OGS (non-relativistically
and relativistically) by studying the motion of a single particle
moving in two spatial dimensions in a specified potential. \ The
motion in this case is readily visualizable, and the different types
of motion can be classified into three categories: annulus, where
each particle always crosses the other two in succession; pretzel,
in which two particles cross each other at least twice in a row before
either crosses the third; and chaotic, where the sequence of particle
crossings does not progress in a discernible pattern.

In this paper we carry out an investigation of the 4-body OGS. Analogous
to its 3-body counterpart, this system is isomorphic to the motion
of a single particle moving in three spatial dimensions in a specified
potential. \
Consequently the 4-body case is of particular interest in that it
is the largest value of $N$ for which the motion of the system can
be directly visualized\textcolor{red}{{} }\textcolor{black}{(we note
that evidence has been provided that when $N\geqq11$ there is no
segementation of the phase space and the system is ergodic} \cite{Reidel}).
\ We consider only the non-relativistic case (which to our knowledge
has never been studied), leaving the relativistic case for future
work.

The outline of our paper is as follows. We begin with an overview
of the problem of 4-body motion in Newtonian gravity, and describe
a general classification \textcolor{black}{scheme for the motion of
the particles that is an extension of that employed in the 3-body
case \cite{Burnell}. \  Two numerical solution methods employed
in the paper are described: the first using numerical integration
and the equations of motion to obtain smooth particle trajectories.
The second method uses collisions between particles as time steps
and maps between the collisions, providing a means to analyze trajectories
and accurately calculate Lyapunov exponents at very large time scales.
Utilizing two solution methods also provided a useful cross check
for results obtained. We then specify to a system of equal masses,
and describe sample trajectories of various dimensionality. \ Following
this we present a proposal for constructing Poincare plots for the
various trajectories encountered in this system. These plots are three-dimensional
generalizations of the two-dimensional plots constructed for the three-body
case \cite{LMiller}. Although difficult to visualize in complete
generality, they do provide interesting information concerning the
chaotic behaviour of the system. We also analyze the Lyapunov exponents
for the system we consider to obtain measures of stochasticity. The
analysis was done using a method due to Benettin et.al.\cite{Benettin}
for calculating the largest Lyapunov exponent. These results are very
consistent with what would be expected from the plots from qualitative
assessment of the stochasticity of the different trajectories. We
find as well an unexpected feature of some Lyapunov graphs where the
perturbed and unperturbed trajectories diverged significantly from
one another. We refer to this effect as orbital bifurcation, and find
that it is caused by small changes in the collision order between
the two trajectories, ultimately leading to large differences in their
qualitative behaviour.} We then close our paper with some concluding
remarks.

\section{Four-Body Motion in Newtonian Gravity}

In (1+1) dimensional Newtonian gravity, the Hamiltonian of our system
of particles is given by:
\begin{equation}
H=\sum_{a=1}^{4}\frac{p_{a}^{2}}{2m_{a}}+\pi G\sum_{a=1}^{4}\sum_{b=1}^{4}m_{a}m_{b}\left|z_{a}-z_{b}\right|\label{newtham}
\end{equation}
which is simply the sum of the kinetic energies of the particles and
the potential interactions between them. The gravitational potential
is determined from 
\begin{equation}
\nabla^{2}\phi=4\pi G\rho=4\pi G\sum_{a=1}^{4}m_{a}\delta(x-z_{a})\label{laplace 1d}
\end{equation}
where $\rho$ is the mass density of the system, here modeled as a
set of four point particles. In one spatial dimension the solution
to this equation is 
\begin{equation}
\phi\left(x\right)=2\pi G\sum_{a=1}^{4}m_{a}\left|x-z_{a}\right|\label{laplacesoln}
\end{equation}
and the potential is given by $V=\frac{1}{2}\sum_{a=1}^{4}m_{a}\phi\left(z_{a}\right)$.\textit{\ }The
equations of motion are given by:
\begin{equation}
\dot{z}_{a}=\frac{\partial H}{\partial p{_{a}}},\qquad\dot{p}_{a}=-\frac{\partial H}{\partial z{_{a}}}\label{newtmot}
\end{equation}

Since the momentum is conserved and since the potential depends only
on the separations between the particles, there are actually only
six \textit{independent} degrees of freedom in the 4-body system:
the three separations between the particles and their conjugate momenta.
\ This is easily seen by making the following convenient change of
coordinates 
\begin{equation}
\begin{array}{cc}
z_{12}=\sqrt{2}\rho & z_{34}=\sqrt{2}\alpha\\
z_{13}=\frac{1}{\sqrt{2}}(\rho+\sqrt{3}\beta-\alpha) & z_{23}=\frac{1}{\sqrt{2}}(-\rho+\sqrt{3}\beta-\alpha)\\
z_{24}=\frac{1}{\sqrt{2}}(-\rho+\sqrt{3}\beta+\alpha) & z_{14}=\frac{1}{\sqrt{2}}(\rho+\sqrt{3}\beta+\alpha)
\end{array}\label{changevar1}
\end{equation}
where $z_{ij}=z_{i}-z_{j}$. \ The conjugate momenta are given by
\begin{equation}
\begin{array}{cc}
p_{1}=\frac{1}{\sqrt{2}}(p_{\rho}+\frac{\sqrt{3}}{2}p_{\beta}) & p_{2}=\frac{1}{\sqrt{2}}(-p_{\rho}+\frac{\sqrt{3}}{2}p_{\beta})\\
p_{3}=\frac{1}{\sqrt{2}}(p_{\alpha}-\frac{\sqrt{3}}{2}p_{\beta}) & p_{4}=\frac{1}{\sqrt{2}}(-p_{\alpha}-\frac{\sqrt{3}}{2}p_{\beta})
\end{array}\label{changevar2}
\end{equation}
where we have set $p_{1}+p_{2}+p_{3}+p_{4}=0$ by conservation of
momentum. \
Without loss of generality the centre of mass of the system can be
fixed at the origin. \textcolor{black}{The reason for this choice
of coordinate transformation is to produce a symmetric potential in
3 spatial dimensions. It can be obtained by requiring that it reduce
to the more familiar 3-body-like potential }\cite{Goodings}\textcolor{black}{{}
when two particles are placed directly on top of one another (i.e.
when one of $z_{12}$, $z_{23}$, or $z_{13}$ vanish).} \bigskip{}

In the equal mass case the Hamiltonian (\ref{newtham}) becomes
\begin{align}
H & =\frac{1}{2m}(p_{\rho}^{2}+p_{\alpha}^{2}+\frac{3}{2}p_{\beta}^{2})+\frac{8\pi Gm^{2}}{\sqrt{8}}\left[\left|\rho\right|+\left|\alpha\right|+\frac{1}{2}\left|\rho+\alpha+\sqrt{3}\beta\right|\right.\notag\\
 & \left.+\frac{1}{2}\left|\rho-\alpha+\sqrt{3}\beta\right|+\frac{1}{2}\left|\rho+\alpha-\sqrt{3}\beta\right|+\frac{1}{2}\left|\rho-\alpha-\sqrt{3}\beta\right|\right]\label{hamnew}
\end{align}
which is the Hamiltonian of a single particle moving in three spatial
dimensions in a linear potential whose shape is that of a 3-simplex.
\ In general a the $N$ particle OGS can be mapped to a single particle
moving in $N-1$ dimensions in a linear potential whose equipotential
surfaces are that of an $N-1$ simplex. \ As such, the 4-body OGS
is the largest system for which the motion can be directly visualized.
\ We shall refer to the form (\ref{hamnew}) as the Hamiltonian of
the box-particle.

\begin{figure}[tbph]
\begin{centering}
\includegraphics[width=1\linewidth]{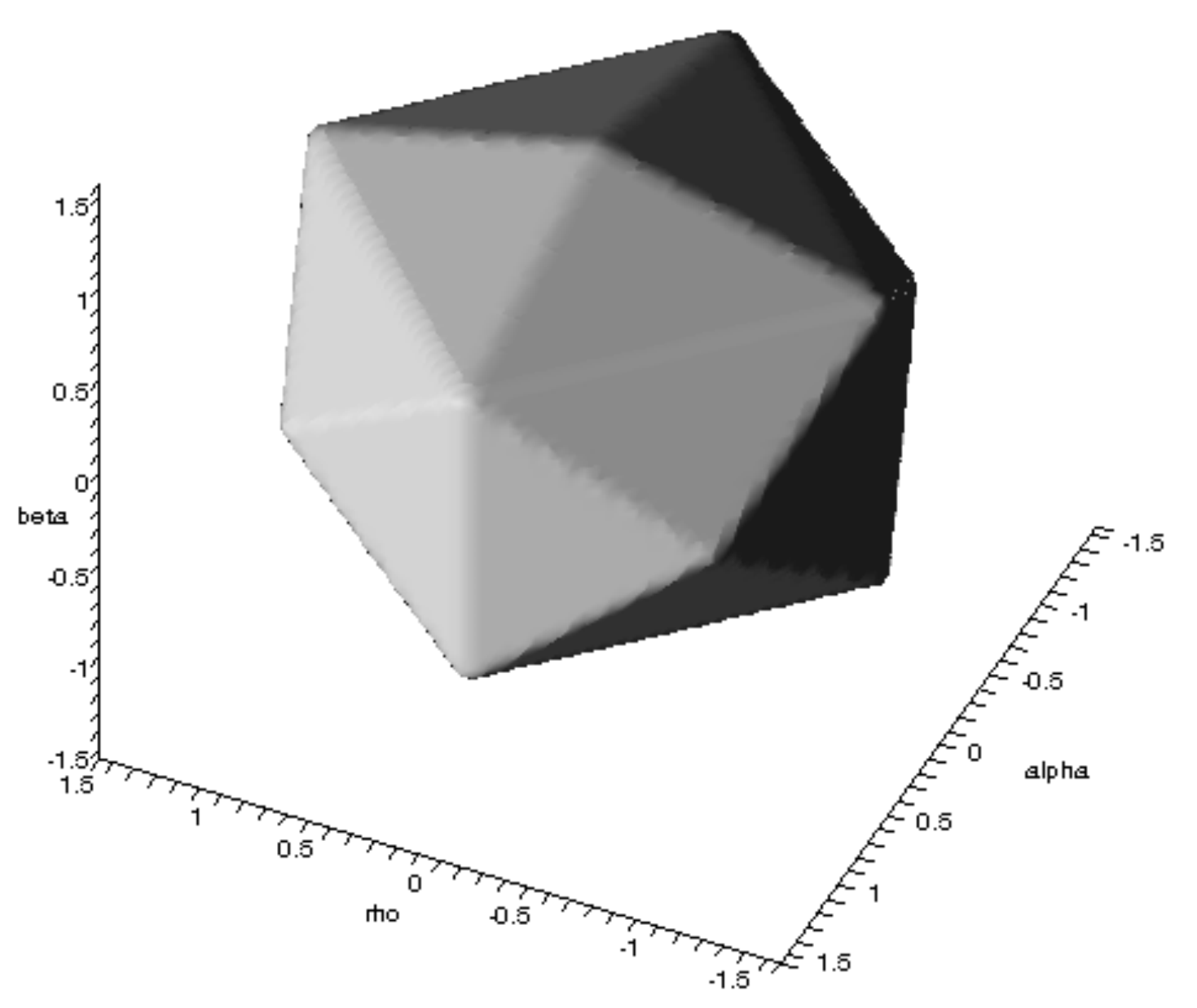}
\par\end{centering}

\caption{Shape of the Newtonian potential in the equal mass case, for a certain
value of $V$. \ Larger values of $V$ will produce larger scaled
``boxes''.}

\label{fig:potential} 
\end{figure}

Using this result, plot the potential in the equal mass case from
equation (\ref{hamnew}), defined as

\begin{equation}
V(\rho,\beta,\alpha)=H(p_{\rho}=0,p_{\beta}=0,p_{\alpha}=0)\label{potential}
\end{equation}
which can be drawn in $(\rho,\beta,\alpha)$\ space as shown in Figure
\ref{fig:potential}. An equipotential surface is that of a cube of
pyramid-shaped sides. \ A cross-section of this surface through any
of the edges of one of these pyramids yields a hexagon whose sides
are not all of equal length. This is reflective of the fact that on
such cross-sections the problem reduces to that of the three-body
problem with unequal masses \cite{justinrobb}\ since two particles
occupy the same position.

\subsection{Methods for Solving the Equations of Motion}

We turn now to the problem of solving the equations of motion (\ref{newtmot}).
As there are no singularities in the potential whenever two particles
cross, we assume that the particles pass through each other freely
upon collision. \ \ Prior to any collision, solving the equations
is trivial: since the acceleration of each particle is constant at
any given instant, the trajectory of each particle is a quadratic
function of time. However after each subsequent crossing the acceleration
of a given particle changes its magnitude, since the number of bodies
to the right and left of it have changed. \ Hence an analytic closed-form
solution to the equations of motion (\ref{newtmot}) is completely
impractical. \ 

We therefore solve the equations of motion numerically, using a Matlab
ODE\
(ordinary differential equation)\ routine to integrate the equations
of motion (\ref{newtmot}). \ For the most part the standard ``ode45''
routine proved to be completely sufficient for our needs. \ In some
cases we also used the ``ode113'' routine because of ou\textcolor{black}{r
extremely stringent error tolerances ($10^{-9}$\ relative and $10^{-10}$\ absolute).
\ The former is based on the Runge-Kutta formula while the latter
is a variable order Adams-Bashforth-Moulton PECE solver (see the Matlab
ODE\ documentation for more detail and references). This solution
method was very useful for mapping the Lissajous figures and particle
trajectories along smooth paths, allowing one to see trajectory patterns,
such as annulus and pretzel, more clearly.}

\textcolor{black}{Integration of the equations of motion is not sufficient
for computation of the largest Lyapunov exponents, and so we employed
a different solution method. The differential equation solver yields
numerical errors that are negligible on the smaller time scales used
for mapping trajectories and Poincare plots. However, when analyzing
Lyapunov exponents, very large time scales are required to obtain
reliable asymptotic behaviour. Numerical errors from the differential
equation solver cause Lyapunov graphs to diverge after only a few
hundred time steps. We therefore compute the trajectories from collision
to collision, a feasible problem that can can be solved in closed
form because all particles follow paths of constant acceleration in
between collisions. Using this method, such numerical divergences
are avoided and stable Lyapunov graphs can be obtained. }

\textcolor{black}{As a cross-check on our methods, we find that the
Lissajous plots are very similar and Poincare plots precisely the
same between the two methods. This is confirmed through careful analysis
of the Lissajous figures. As will be seen, the dimensions of the figures
are the same and the features, such as the bands and stripes, seen
clearly on the collision method plots, are shared between plots from
the two methods. Furthermore, the Poincare plots can be matched point-to-point
between the two methods, further confirming that the solution methods
are equivalent (these are not shown for simple reasons of redundancy).}

\textcolor{black}{We employ two methods of analysis. \ One is that
of plotting the trajectories of the box-particle in $(\rho,\beta,\alpha)$\ space
for a variety of initial conditions. We also plot the motions of the
four particles as a function of time for each case. \ This provides
an alternate means of visualizing the difference between various types
of motions that can arise in the system.}

To perform the numerical analysis we rescale the variables to have
dimensionless values
\begin{equation}
p_{i}=M_{tot}c\hat{p_{i}}\qquad z_{i}=\frac{4}{\kappa M_{tot}c^{2}}\hat{z_{i}}\label{unitrescale1}
\end{equation}
where $M_{tot}=4m$ is the total mass of the system and $\hat{p_{i}}$
and $\hat{z_{i}}$ are the dimensionless momenta and positions respectively.
\ The dimensionless mass and Hamiltonian are:
\begin{equation}
\eta=\frac{H}{M_{tot}c^{2}}\qquad\hat{m_{i}}=\frac{m_{i}}{M_{tot}}\label{unitrescale2}
\end{equation}
Using the dimensionless variables in (\ref{unitrescale1}) and (\ref{unitrescale2}),
the Hamiltonian (\ref{newtham}) becomes:
\begin{equation}
\eta=\sum\limits _{a=1}^{4}\frac{\hat{p_{a}^{2}}}{2\hat{m_{a}}}+\frac{1}{2}\sum\limits _{a=1}^{4}\sum\limits _{b=1}^{4}\hat{m_{a}}\hat{m_{b}}\left|\hat{z_{a}}-\hat{z_{b}}\right|\label{hamrescale}
\end{equation}
As for the equations of motion (\ref{newtmot}) one gets:
\begin{eqnarray}
\frac{\partial\eta}{\partial\hat{p_{i}}} & = & \frac{d\hat{z}_{i}}{d\hat{t}}\label{motrescale1}\\
\frac{\partial\eta}{\partial\hat{z_{i}}} & = & -\frac{d\hat{p_{i}}}{d\hat{t}}\label{motrescale2}
\end{eqnarray}
where $t=\frac{4c}{8\pi GM_{tot}}\hat{t}$. \ 

$ $

A time step in the numerical code has a value $\hat{t}=1$. \ All
the numerical calculations were carried out using the rescaled variables
(\ref{unitrescale1}). \ Henceforth we drop all of the ``hats''
of the dimensionless variables for convenience.

Note that since the energy is a constant of the motion, we could further
rescale all quantities in terms of $\eta$, thereby fixing this final
redundant scale. However we shall find it convenient to employ the
above rescalings, as it affords us more freedom in choosing initial
conditions.

\subsection{Classifying the Motions}

Prior to any collision between the particles the evolution of the
system is straightforward: each particle moves with a constant acceleration
that is proportional to the difference between the total mass on its
right and left sides. \ However after a collision, where we assume
that the particles pass through each other, the mass difference changes,
and with it the accelerations of the particles. \ It is these repeated
changes in the accelerations of the particles that yield the interesting
dynamics of the system. \ 

From the perspective of the box particle, such crossings correspond
to the box particle crossing any plane that bisects the 3-simplex
through its vertices and edges, yielding a discontinuous change in
the box particle's acceleration. These planes occur in pairs whose
line of intersection is along each of the three principal axes, for
a total of six such planes. Their equations are given by setting any
one of the six quantities in eq. (\ref{changevar1}) to zero, and
each plane corresponds to the crossing of a pair of particles. For
example $\rho=0$\ corresponds to the crossing of particles 1 and
2, whereas $\rho=\sqrt{3}\beta+\alpha$\ corresponds to the crossing
of particles 2 and 4. \ 

Initially we have all of the crossings listed in order as a string
of vanishing quantities. $\{z_{12}z_{13}z_{14}z_{23}z_{24}z_{34}\}$.
The direction of crossing is irrelevant; for example 1 crossing 2
is equivalent to 2 crossing 1. \ At any given instant we can fix
the positions of particles in a certain order from left to right,
i.e. $(1,2,3,4)$, in which case we only have 3 possible crossings:\ $\{z_{12},z_{23},z_{34}\}$.
\ We denote these using the Braid operators $\{\sigma_{1},\sigma_{2},\sigma_{3}\}$,
with $\sigma_{1}$\ corresponding to an interchange between the right-most
pair of particles, $\sigma_{2}$\ for the middle pair and $\sigma_{3}$\ for
the left-most pair. More generally, instead of defining the actual
particles as $(1,2,3,...)$, we define the positions as that sequence:
the left-most particle is at position $1$, next $2$, and so on with
the right-most particle being at position $N$. Given any sequence
of particle crossings, we can employ Braid Group notation \cite{braidgroup,braidword}
to classify the motion, denoting pair crossings with the set $\{\sigma_{1},\sigma_{2},\ldots,\sigma_{N-1}\}$.
For example the sequence $z_{12}z_{13}z_{23}z_{14}z_{24}...$\ is
described by $\sigma_{1}\sigma_{2}\sigma_{1}\sigma_{3}\sigma_{2}$,
and the initial configuration $(1,2,3,4)$\ becomes the final configuration
$(3,4,2,1)$. \ Since crossing direction is irrelevant, the reciprocal
notation for the Braid Group is ignored (i.e. $\sigma_{j}=\sigma_{j}^{-1}$\ \ in
this context). \textcolor{black}{We do not concern ourselves with
the permutation properties of the braid groups. We use this notation
out of convenience in classifying the sequence of motion of the particles.
Since the particular sequence of collision is important, any permutation
of the operators would result in loss of information about the motion
in the system. \ }

\bigskip{}

We are now ready to classify the distinct kinds of motion that can
occur. Consider first the 3-body case. This problem can be mapped
to that of a single particle moving in a hexagonal-shaped well, with
the bisectors of the hexagon denoting pair-crossings of the particles
\cite{Burnell}. In this case we have $\{\sigma_{1},\sigma_{2}\}$\ as
the Braid operators. \ For any string of these we can classify the
motion in the equal mass case as follows:

\begin{equation}
\begin{array}{cc}
\sigma_{1}\sigma_{1}\;,\;\sigma_{2}\sigma_{2} & \text{A motion}\\
\sigma_{1}\sigma_{2}\;,\;\sigma_{2}\sigma_{1} & \text{B motion}
\end{array}\label{3bodymotions}
\end{equation}
by comparing subsequent items in the string; we have included the
$A/B$ descriptors employed previously in 3-body studies \cite{Burnell}.
\ In other words, since crossing direction is irrelevant, the only
interesting types of motion are when the same pair of particles crosses
twice in a row ($A$-motion) corresponding to crossing a single bisector
of the hexagon twice in succession, or when one particle crosses each
of its compatriots in succession ($B$-motion) corresponding to the
crossing of two successive bisectors. \ In the unequal mass case
the hexagon is no longer symmetric, and so $\sigma_{1}\sigma_{1}$,
say, is now distinguishable from $\sigma_{2}\sigma_{2}$. \ However
one could impose an equivalence relation between these two motions
and continue to employ the above notation. \ Any given motion in
the system can be characterized by a sequence of letters $A$\ and
$B$\ (called a symbol sequence), with a finite exponent $n$\ denoting
$n$-repeats and an over-bar denoting an infinite repeated sequence.

This idea extends naturally to the 4-body system. Here the Braid operators
are $\{\sigma_{1},\sigma_{2},\sigma_{3}\}$. \ We can construct the
following definitions:

\begin{equation}
\begin{array}{cc}
\sigma_{1}\sigma_{1}\;,\;\sigma_{2}\sigma_{2}\;,\;\sigma_{3}\sigma_{3} & \text{A motion}\\
\sigma_{1}\sigma_{2}\;,\;\sigma_{2}\sigma_{1}\;,\;\sigma_{2}\sigma_{3}\;,\;\sigma_{3}\sigma_{2} & \text{B motion}\\
\sigma_{1}\sigma_{3}\;,\;\sigma_{3}\sigma_{1} & \text{C motion}
\end{array}\label{4bodymotions}
\end{equation}
\begin{figure}[tbph]
\begin{centering}
\includegraphics[width=1\linewidth]{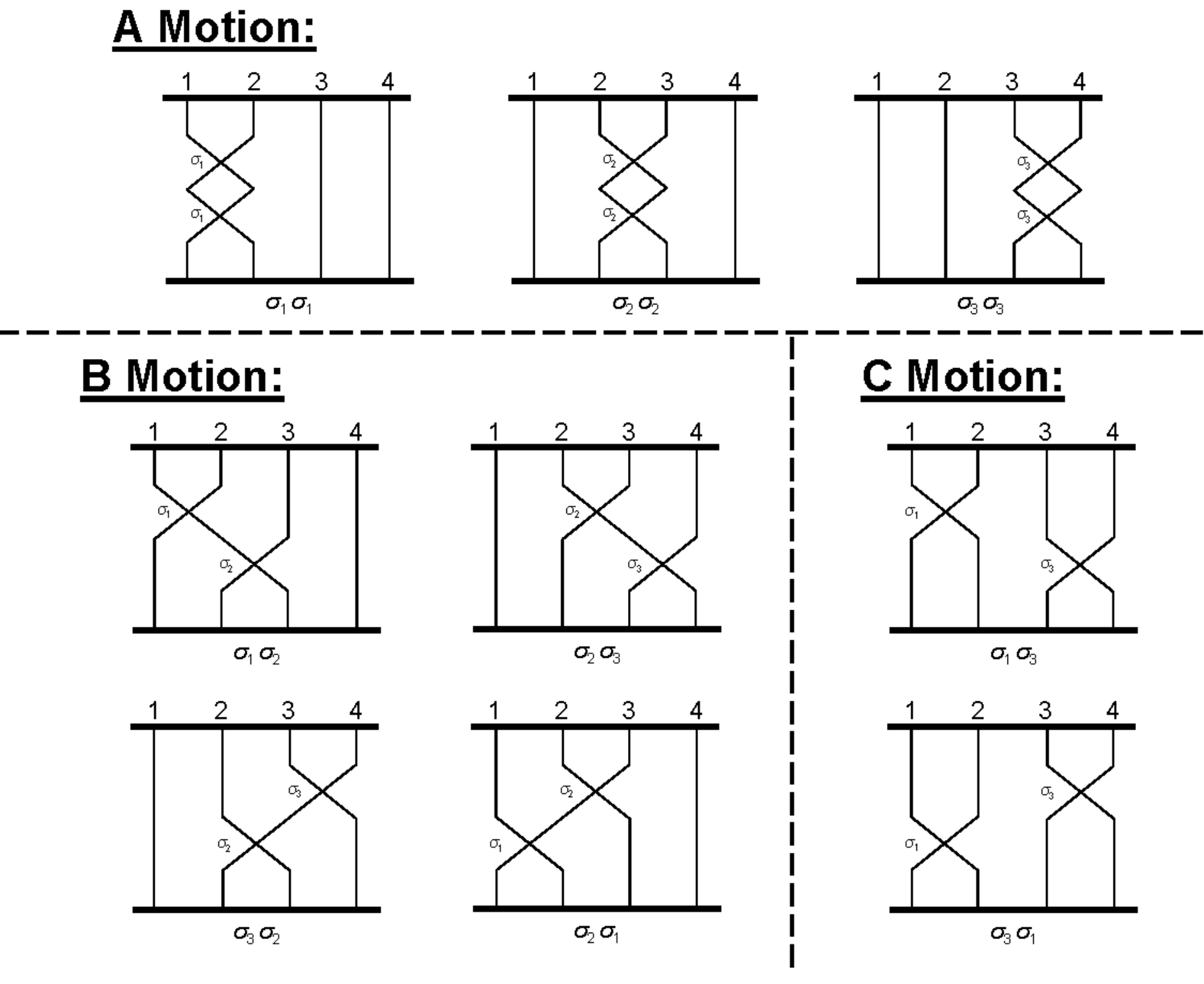}
\par\end{centering}

\caption{4-body motion classes.}

\label{fig:4body_motion_class} 
\end{figure}
whose results can be visualized in Figure \ref{fig:4body_motion_class}.
The $A$\ and $B$\ motions still represent the same physical situations
as in the 3-body case. However there is now a new type of motion:
$C$\ motion, which is when two particles cross one another and then
the other two cross one another.

We can proceed further, generalizing our arguments and definitions
to the $N$-body case and make a more formal definition of our motion
classes whose specific cases for $N=3$\ and $N=4$\ will be equivalent
to what we have described above.

To describe crossings of $N$\ particles for a specific trajectory
we have the Braid operators $\{\sigma_{1},\sigma_{2},\ldots,\sigma_{N}\}$.
\ A sequence of $m$\ pair crossings will be described by

\begin{equation}
\sigma_{f(1)}\sigma_{f(2)}\sigma_{f(3)}...\sigma_{f(m)}\label{braidworddef}
\end{equation}
where $1\leq f(x)\leq(N-1)$\ for all $1\leq x\leq m$\ is a discrete
integer function\ . \ Here\ $\sigma_{f(x)}$\ means that the particles
currently in positions $f(x)$\ and $f(x)+1$\ cross. \ Any given
sequence of $m$\ Braid operators forms a unique ordered list of
crossings for the given trajectory. \ As stated previously, crossing
directions are irrelevant.

Now we define a new function

\begin{equation}
g(x)\equiv|\Delta f(x)|=|f(x+1)-f(x)|\label{braiddiffdef}
\end{equation}
using the finite forward difference function of $f$\ \cite{forwarddiff}.
We are interested in the absolute difference between subsequent terms,
and thus we have $0\leq g(x)\leq(N-2)$\ for all $1\leq x\leq(m-1).$\ Now
$g(x)$\
defines a metric that describes the relative ``distance'' between
any pair of crossings, and we classify the motion according to this
distance \ 

\begin{equation}
\begin{array}{cc}
g(x) & \text{Motion Class}\\
0 & A\\
1 & B\\
2 & C\\
..
\end{array}\label{braiddiffmotions}
\end{equation}
denoting each type by increasing letters of the alphabet. In other
words, $A$-motion corresponds to any 2 crossings in nearest proximity
-- two particles cross each other twice in succession. \ $B$-motion
corresponds to any 2 crossings in next-nearest proximity -- two particles
cross each other, and then one of them crosses its other nearest neighbour.
\ $C$-motion corresponds to any 2 crossings in next-to-next-nearest
proximity: two particles cross each other and then a neighbouring
pair cross each other. \
We can continue on in this fashion until we reach the extreme case
in which the right-most pair of particles cross one another followed
by the crossing of the left-most pair (or vice-versa).

We illustrate this classification with some examples. For $N=3$,
we have $1\leq f(x)\leq2$. \ Suppose we have a crossing sequence
$\sigma_{1}\sigma_{2}\sigma_{1}\sigma_{1}\sigma_{2}$, yielding from
(\ref{3bodymotions}) the symbol sequence $BBAB$. By our definition
above we have

\begin{equation}
\begin{array}{cccccc}
x & 1 & 2 & 3 & 4 & 5\\
f(x) & 1 & 2 & 1 & 1 & 2\\
g(x) & 1 & 1 & 0 & 1
\end{array}\label{3bodyexamplef}
\end{equation}
and from (\ref{braiddiffmotions}) we get $BBAB$\ as we expected.
\ For $N=4$, we have $1\leq f(x)\leq3$. \ Consider the previous
example $\sigma_{1}\sigma_{2}\sigma_{1}\sigma_{3}\sigma_{2}$, which
from (\ref{4bodymotions}) yields the symbol sequence $BBCB$, the
same result we would obtain from computing successive values of $g(x)$\ (which
are $1,1,2,1$\ for this example).

The preceding classification system is limited to pair-wise crossings
and does not cover situations in which more than one pair of particles
crosses at the exact same time step. \ While the braid group notation
does allow for multiple collisions by writing them ``left-to-right'',
it is important in our system whether or not these collisions occur
at the same time step.\ \
For the $N$-body problem, a simultaneous collision of $m$\ particles
corresponds to the crossing of a single particle through an $(N-m)$-dimensional
surface in the interior of the $(N-1)$\ simplex. \ This surface
is obtained by continually bisecting the simplex along its (higher-dimesional)
edges and vertices until the surface of appropriate dimensionality
is obtained. \ We can denote such collisions by extending the braid
group notation with the set $\{\sigma_{1^{m}},\sigma_{2^{m}},\ldots,\sigma_{(N+1-m)^{m}}\}$,
where the subscript denotes which set of particles is involved, beginning
with the left-most, and the superscript (on these subscripts) denotes
the number of particles in the collision. \ For example $\sigma_{9^{7}}$\ denotes
a 7-particle collision that involves particles 9-15. \ We shall drop
the superscript ``2'' when pair-wise collisions are involved. All
collisions yield crossings except for the situation in which the initial
conditions cause $m$\ particles to occupy the same point throughout
the motion. \ In this latter case the system reduces to that of an
(unequal mass) $\left(N-m\right)$-body problem.

Of course for the 3 and 4-body cases the numbers of multiple collisions
are simple. There is a single type of multiple collision in the 3-body
case, which occurs when the hex particle crosses the origin. \ In
the 4-body case we can have two kinds of 3-body collisions (described
by \ $\{\sigma_{1^{3}},\sigma_{2^{3}}\}$) that occur when the box
particle crosses the line of intersection of any two bisecting planes
of the 3-simplexes. \ We also have two kinds of 4-body collisions\ (described
by $\{\sigma_{1^{2}3^{2}},\sigma_{1^{4}}\}$). \ The former occurs
when two pairs of particles cross each other at the same time, and
corresponds to the box particle crossing one of the three lines connecting
opposite vertices of the pyramids in the simplex\ (see Figure \ref{fig:potential}).
\ The latter is when all four particles cross at once, equivalent
to the\ box particle crossing the origin.

\textcolor{black}{One interesting feature of multiple particle collisions
is that one can always predict the new order of particles given the
preceding order of particles alone so long as all particles cross
simultaneously. Suppose we have a multiple collision of $n$ particles.
Consider two adjacent particles in this multiple collision some small
time just before the collision occurs; we assign the `right' direction
as postive for the purpose of assigning velocities to the particles.
In order for these two particles to cross one another, the particle
on the left must have a larger velocity than the particle on the right,
or else the right particle will be moving away, not towards, the left
particle and no collision would occur. Apply this reasoning to every
adjacent pair and you discern that the left-most particle must have
the largest velocity, decreasing as you move rightward in the sequence
of particles just before the multiple collision. Therefore, the order
immediately after the collision occurs will be the reverse of the
original order because the previously left-most particle will be travelling
rightward faster than all other particles in the collision and emerge
afterwards as the right-most, and so on for all particles in the collision.
Note that if any one of the $n$ particles does not satisfy the increasing
velocity condition then the multiple collision has fewer than $n$
particles.}

\section{Equal Mass Trajectories}

In this section we consider the equal-mass case. We study the behaviour
of the four-body system using a variety of initial conditions, and
compare to the 3-body case \cite{Burnell} where relevant. \ Many
patterns of motion in the 3-body system have natural counterparts
in the 4-body case. \

\subsection{Two-Dimensional Trajectories}

A necessary check on the code is to see if we can reproduce results
in the 3-body case. \ Indeed, if \textit{one} of the box-particle's
position \textit{and} momentum coordinates is initially zero it will
remain zero throughout the motion, and the motion of the box-particle
will be restricted to a plane. This corresponds to the situation mentioned
near the end of the previous section, as is easily shown. From the
Hamiltonian (\ref{hamnew}), the equations of motion for the $\alpha$
coordinate%
\footnote{The choice of $\alpha$ is completely arbitrary; we could have chosen
$\beta$ or $\rho$ equivalently.%
} are:
\begin{align}
\dot{p}_{\alpha} & =-\frac{\partial H}{\partial\alpha}=\frac{8\pi Gm^{2}}{\sqrt{8}}\left[-\frac{1}{2}sgn(\rho-\alpha+\sqrt{3}\beta)-\frac{1}{2}sgn(-\rho-\alpha+\sqrt{3}\beta)\right.\notag\\
 & \left.+\frac{1}{2}sgn(-\rho+\alpha+\sqrt{3}\beta)+\frac{1}{2}sgn(\rho+\alpha+\sqrt{3}\beta)+\frac{1}{2}sgn(\alpha)\right]\label{padot}
\end{align}
\begin{equation}
\dot{\alpha}=\frac{\partial H}{\partial p_{\alpha}}=\frac{p_{\alpha}}{m}\label{adot}
\end{equation}
So if the initial conditions are $\alpha=0$ and $p_{\alpha}=0$,
then we have clearly from eq.(\ref{adot}) that $\alpha=0\;\;$for
all $t$, which implies that eq.(\ref{padot}) becomes
\begin{eqnarray}
\dot{p}_{\alpha} & = & \frac{\kappa m^{2}c^{4}}{\sqrt{8}}\left[-\frac{1}{2}sgn(\rho+\sqrt{3}\beta)-\frac{1}{2}sgn(-\rho+\sqrt{3}\beta)+\frac{1}{2}sgn(-\rho+\sqrt{3}\beta)+\frac{1}{2}sgn(\rho+\sqrt{3}\beta)\right]\notag\\
 & = & 0\label{padot2}
\end{eqnarray}
and so the motion is restricted to the $\left(\rho,\beta\right)$
plane. Similarly, it is not difficult to convince oneself that the
box-particle is also restricted to the $(\alpha,\beta)$ plane when
the initial momentum and position $p_{\rho}$ and $\rho$ are both
zero, and to the $(\alpha,\rho)$ plane when $p_{\beta}$ and $\beta$
are initially both zero. \
Note, however, that even if there is no momentum along $\alpha$ initially,
it does not mean that the particles will never move in the $\alpha$
direction. \ Indeed if the initial position $\alpha_{o}\neq0$, then
the particle will acquire momentum according to eq.(\ref{padot}).

\bigskip{}

\begin{figure}[tbph]
\begin{centering}
\includegraphics[width=1\linewidth]{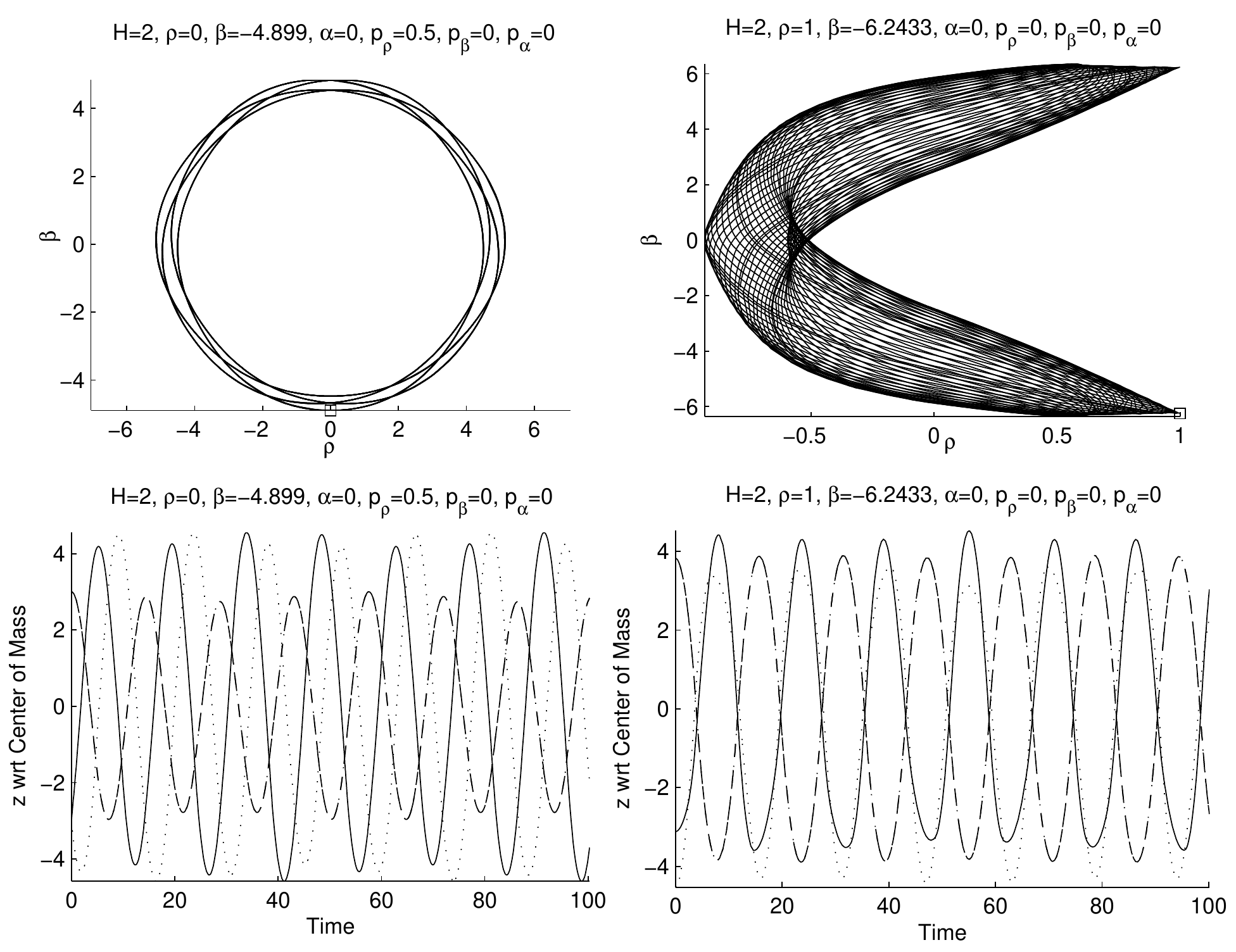}
\par\end{centering}

\caption{Annulus (on the left) and pretzel (on the right) orbits for 500 time
steps and $H=2$. \ The initial conditions for the plots on the left
are: $\rho=0,$ $\alpha=0,$ $p_{\rho}=0.5,$ $p_{\beta}=0,$ $p_{\alpha}=0$.
\ $\beta$ will be calculated so that eq. (\ref{hamnew}) is satisfied
initially. \ The initial conditions for the plots on the right are:\ $\rho=1,$
$\alpha=0,$ $p_{\rho}=0,$ $p_{\beta}=0,$ $p_{\alpha}=0$.}

\label{fig:beg_2d} 
\end{figure}

Consequently we should recover all of the patterns of motion that
the hex particle exhibits in the 3-body case \cite{Burnell} for the
subset of initial conditions in which \ $\alpha=0$ and $p_{\alpha}=0$.
We found this to be the case, and recovered the annulus, pretzel and
chaotic motions referred to in the introduction. \ Figure \ref{fig:beg_2d}
illustrates some examples belonging to the first and second classes.
\textbf{\ }Note that this is not an equal-mass 3-body system, but
rather one with unequal masses, because the two particles that are
moving together are like one single particle with a mass twice as
large. Indeed, two trajectories (out of four) are exactly identical,
which means that two particles are moving together.

Similarly, it is not surprising that we recover the two-body case
if we choose \textit{two} momentum and position coordinates to be
zero initially. \ For example, if we set $\rho_{0}=0$, $\ \alpha_{0}=0$
and $p_{\rho0}=0$, $\ p_{\alpha0}=0$, then the box-particle will
move on a line parallel to the $\beta$-axis. \ In this case, the
$z(t)$ plots show only two distinguishable trajectories, because
we have actually two pairs of particles moving. \ Unlike the reduction
to the three-body case from the equal mass four-body problem, we have
here a reduction to an equal mass two-body problem.

\subsection{Three-Dimensional Trajectories}

\begin{figure}[tbph]
\begin{centering}
\includegraphics[width=1\linewidth]{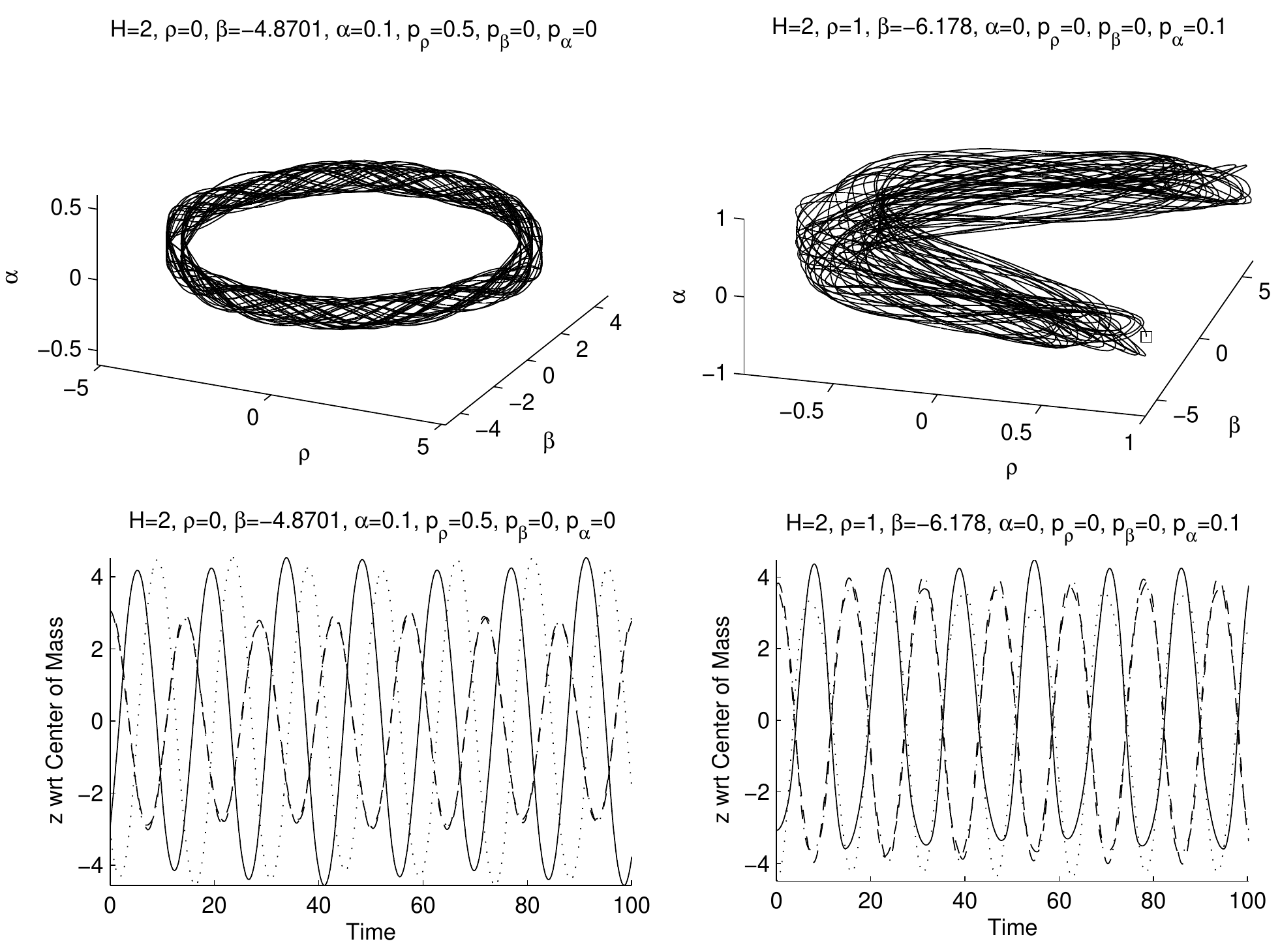}
\par\end{centering}

\caption{Annulus and pretzel orbits for 500 time steps and $H=2$. \ The initial
conditions for the plots on the left are: $\rho=0,$ $\alpha=0.1,$
$p_{\rho}=0.5,$ $p_{\beta}=0,$ $p_{\alpha}=0$.\textcolor{black}{{}
The Lyapunov exponent for this trajectory was calculated to be 1.214$\times$$10^{-2}$.
The initial conditions for the plots on the right are:\ $\rho=1,$
$\alpha=0,$ $p_{\rho}=0,$ $p_{\beta}=0,$ $p_{\alpha}=0.1$. The
Lyapunov exponent for this trajectory was calculated to be 7.350$\times$$10^{-3}$.}}

\label{fig:beg_3d} 
\end{figure}

Of course the more interesting situation is when there is motion in
all spatial directions, generating three-dimensional patterns. \ Changing
the initial conditions a little bit from those used for the plots
shown in Figure \ref{fig:beg_2d} by giving, for example, a small
momentum along the $\alpha$ direction gives us the three-dimensional
trajectories (Figure \ref{fig:beg_3d}). Not surprisingly, these patterns
are direct generalizations of what we observed in Figure \ref{fig:beg_2d},
since the motion in the $\alpha$ direction simply perturbs the patterns
previously obtained in the $\left(\alpha,\beta\right)$ plane when
we set $\alpha=0$ and $\ p_{\alpha}=0$, initially. \ Essentially
the original hex-particle patterns develop a ``thickness'' in the
$\alpha$ direction.

For the three-dimensional annulus, we\ chose initial conditions to
show that a non-zero $\alpha_{0}$ induces a non-zero $p_{\alpha}(t)$,
even if $p_{\alpha0}=0$. \ It is also interesting to compare the
peaks of Figure \ref{fig:beg_2d} with those of Figure \ref{fig:beg_3d}.
\ We see that the three-dimensional trajectories experience a small
deviation (near the peak) in the trajectories of the two particles
that were exactly the same for the two-dimensional box-particle trajectories.
\ The more we increase the value of the initial conditions $\alpha_{0}$
or $p_{\alpha0}$ the bigger the deviation, and in the $(\rho,\beta,\alpha)$
space, the particle will have a larger amplitude in the $\alpha$
direction.

\begin{figure}[tbph]
\begin{centering}
\textbf{\vspace{-3.5cm}
}
\par\end{centering}

\begin{centering}
\includegraphics[width=1\linewidth,height=0.75\linewidth]{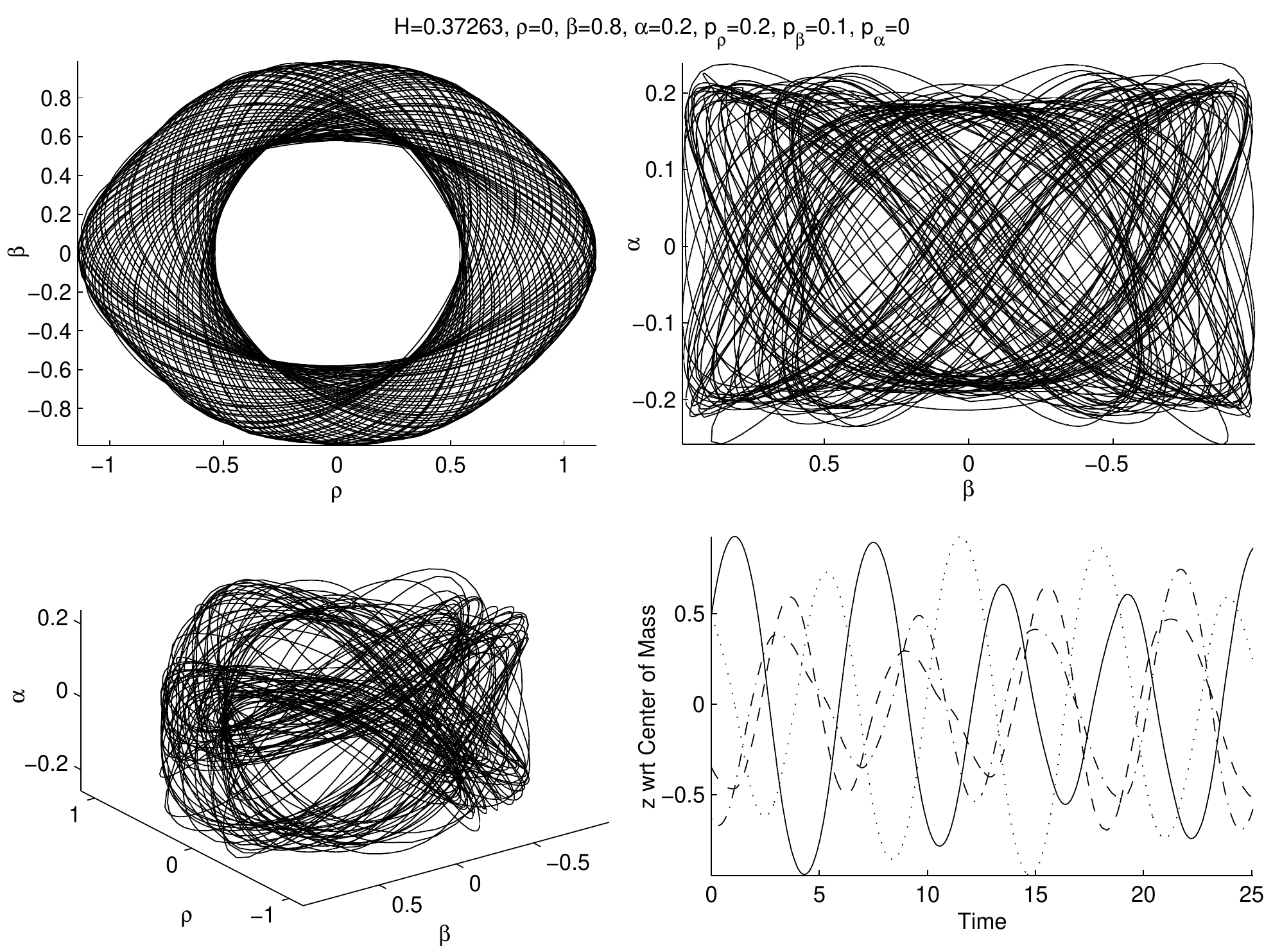}
\par\end{centering}

\begin{centering}
\includegraphics[bb=190bp 0bp 780bp 430bp,clip,width=0.33\linewidth,height=0.45\linewidth]{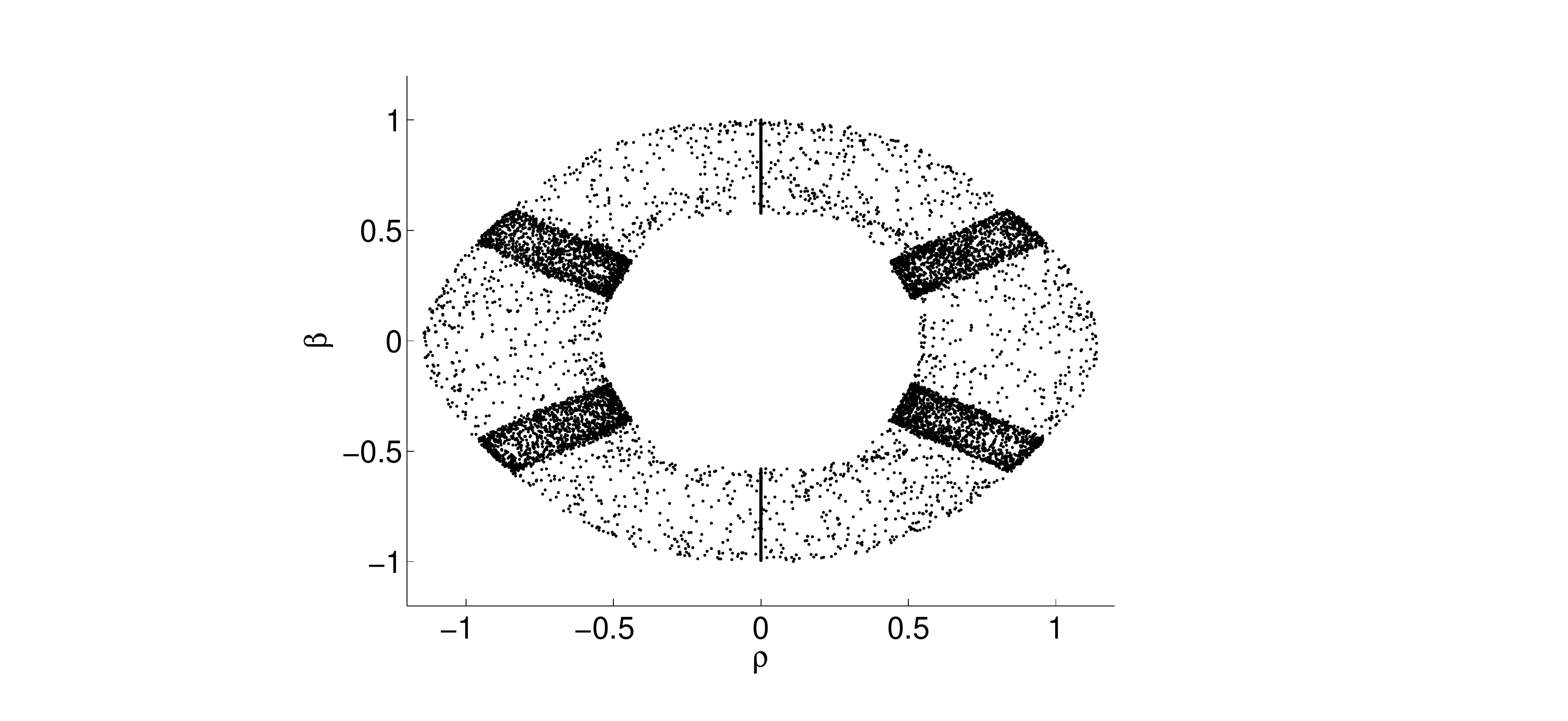}\textcolor{red}{\includegraphics[bb=0bp 100bp 1024bp 471bp,width=0.66\linewidth]{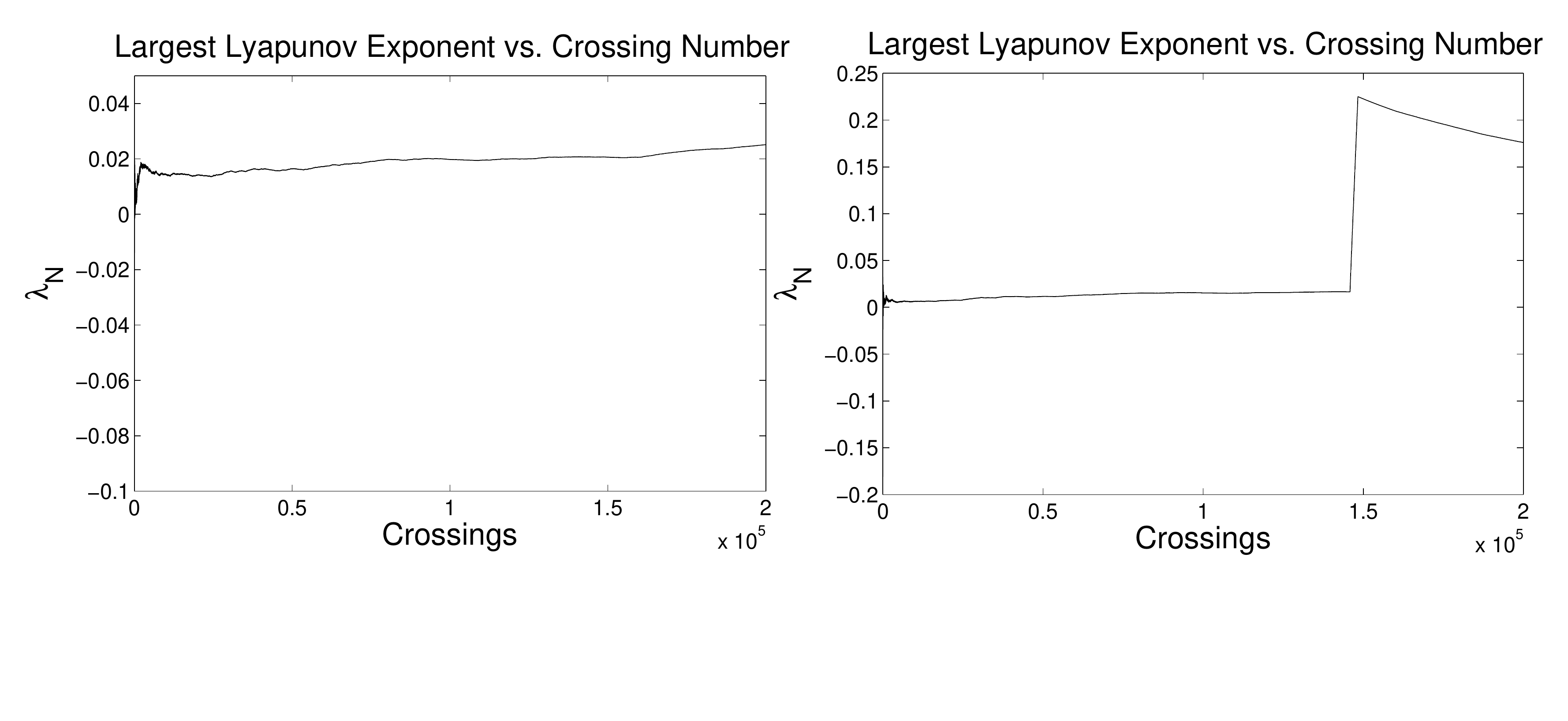}}
\par\end{centering}

\caption{The top two plots show the trajectory from two different directions.
\ It is clear that while being a nice annulus shape in the $(\rho,\beta)$
plane, the plot is much more random in the $(\beta,\alpha)$ plane,
although one can still observe a semi-periodic pattern. \ The bottom-left
isometric plot of the trajectory shows that the periods of oscillation
in the various axes do not line up nicely for these initial conditions.
\ The bottom-right plot show the four particle positions vs. time\textbf{;
}the symbol sequence is\textbf{\ }$\overline{CB^{2}CB^{2}CB^{6}}$.
\textcolor{black}{The bottom-left figure was obtained using the collision
to collision mapping solution. Note the same size, shape and banding
structure between the two figures. -- these are faint in the diagram
at the upper left, but are visible. The Lyapunov graph is shown at
the bottom-middle with a calculated Lyapunov exponent of 2.138$\times$$10^{-2}$.
The Lyapunov graph on the bottom-right is an example of orbital bifurcation
where the trajectories reconverge after some number of collision steps.
It is also seen that even in the 'fixed' trajectory, the Lyapunov
graph does not clearly converge. This seems likely due to problem
within the numerics of the system.}}

\label{fig:ann_mess} 
\end{figure}

\begin{figure}[tbph]
\begin{centering}
\includegraphics[width=1\linewidth]{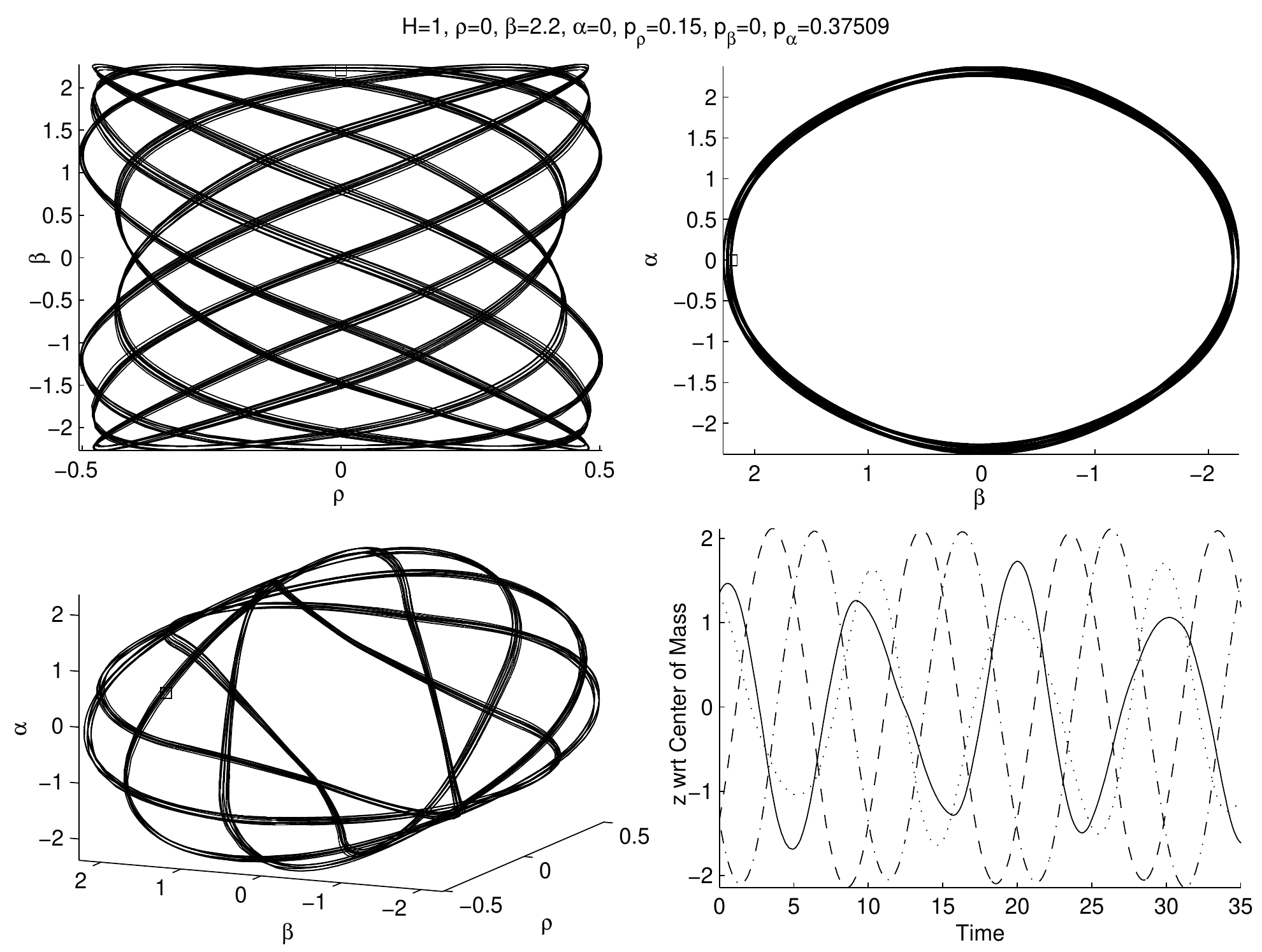}
\par\end{centering}

\begin{centering}
\includegraphics[width=1\linewidth]{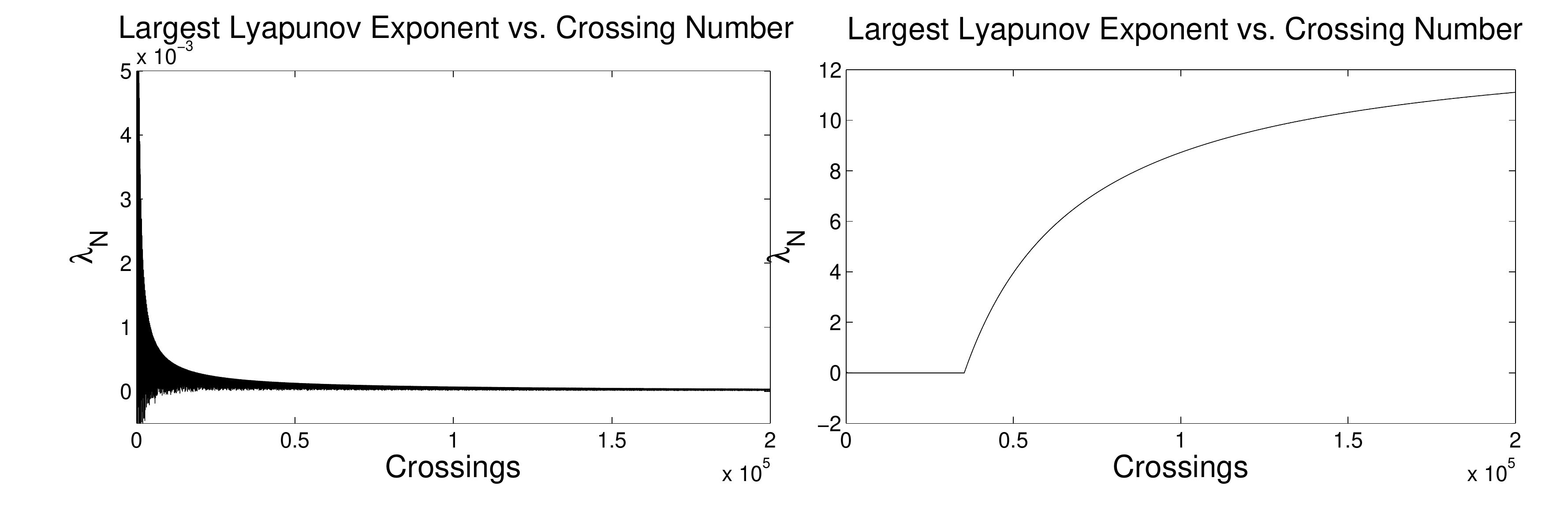}
\par\end{centering}

\caption{The layout of the plots is the same as Figure \ref{fig:ann_mess}.
\
In this case however, we see that from every direction the trajectory
forms a recognizable two-dimensional plot (third direction is not
shown, but it is a pretzel similar to the top-left). \ Since the
periods of the motions in various axes are compatible in this case,
we see from the isometric plot (middle-left) that a fully three-dimensional
periodic trajectory exists. \
Note that FE ($H=1$)\ constraints were used in this case for conve\textcolor{black}{nience.
At the lower-left is the Lyapunov graph for this trajectory, whose
Lyapunov exponent was calculated to be 2$.835\times10^{-5}$ . The
figure at the lower right is another example of orbital bifurcation.}}

\label{fig:ann_pretz} 
\end{figure}

To further investigate the trajectories that can be obtained in the
full three-dimensional case, we follow the method in \cite{Burnell}
- that is using initial condition constraints of \textit{fixed-energy}
(FE)\ and of \textit{fixed-momentum} (FM).\textcolor{black}{{} Since
the Hamiltonian is a homogeneous function of the coordinates and momenta,
it can always be rescaled to unity by an appropriate rescaling of
the phase space variables. Fixed energy conditions are equivalent
to rescaling all variables in terms of $\eta$, as noted previously.
Anticipating future comparison with the relativistic case, we find
it convenient to choose different }initial values for $H,\rho,\beta,\alpha,p_{\rho},$
and $p_{\beta}$, adjusting $p_{\alpha}$ using the Hamiltonian constraint
(\ref{hamnew}), and check that $H$ remains at its initial value
throughout the motion (Figures \ref{fig:beg_2d}, \ref{fig:beg_3d},
\ref{fig:ann_pretz}, \ref{fig:2d_pretz_poinc}, \ref{fig:odd_shape},
\ref{fig:nice_poinc}). \ We have chosen to vary $p_{\alpha}$, to
more easily facilitate comparison with the previous results in the
equal-mass 3-body case \cite{Burnell}.\ \ Similarly for the fixed-momenta
case \textcolor{black}{(Figure \ref{fig:ann_mess}, Figure \ref{fig:1_1_1_1}
- \ref{fig:1_1_2}) }, we will choose $\rho,\beta,\alpha,p_{\rho},p_{\beta}$
and $p_{\alpha}$ while allowing $H$ to vary as the initial conditions
vary. \textcolor{black}{This allows us to more easily select qualitatively
different types of motions. }

We imposed error tolerances on the Matlab ODE\ routine and checked
that the total energy of the system remained constant throughout the
motion for any given set of initial conditions. \ Since we were using
an improved version of the code, we were able to get the error down
to $10^{-8}$ in most cases, and always less than $10^{-7}$. \ 

As noted above, small perturbations of the three-body case yield three-dimensional
trajectories that usually have a nice shape similar to that of the
three-body case if projected onto one of the planes $(\rho,\beta),(\rho,\alpha)$
or $(\beta,\alpha)$. \ However, the third axis is generally just
a periodic oscillation with no real pattern relative to the other
axes, as in Figure \ref{fig:ann_mess}. \ This trajectory's motion
is characterized by the symbol sequence $\overline{CB^{2}CB^{2}CB^{6}}$.
\textcolor{black}{The Lyapunov graph shown in the bottom-middle is
the `fixed' Lyapunov graph where the perturbation was sufficiently
small to prevent orbital bifurcation. The bottom right figure displays
orbital bifurcation where, after some number of collision steps, the
trajectories reconverge suddenly. This is seen by the sudden curve
downwards. In theory, if this trajectory was run for a sufficiently
large number of collision steps, the value would converge to the Lyapunov
exponent of the `fixed' trajectory.}

However, cases where these periods do line up can be obtained with
carefully chosen initial conditions. \ For example, figure \ref{fig:ann_pretz}
shows a trajectory that has a pretzel form when projected onto two
of the planes ($(\rho,\beta)$\ and $(\rho,\alpha)$), and an annulus
when projected onto the third ($(\beta,\alpha)$). \ The three-dimensional
isometric (in which there is no perspective so that objects further
away do \textit{not }appear smaller) view shows a periodic three-dimensional
path - a qualitatively new feature that has no analogue in the 3-body
system. \ The motion here is described by $\overline{(CB^{2}CB^{2}CB^{6})^{2}CB^{6}}$,
which is very similar \textcolor{black}{to the previous figure. This
trajectory also displays orbital bifurcation in its Lyapunov graph
(shown bottom-right). In this case, the trajectories do not reconverge
in the 200,000 collision steps in this simulation.}

\begin{figure}[tbph]
\begin{centering}
\includegraphics[width=1\linewidth]{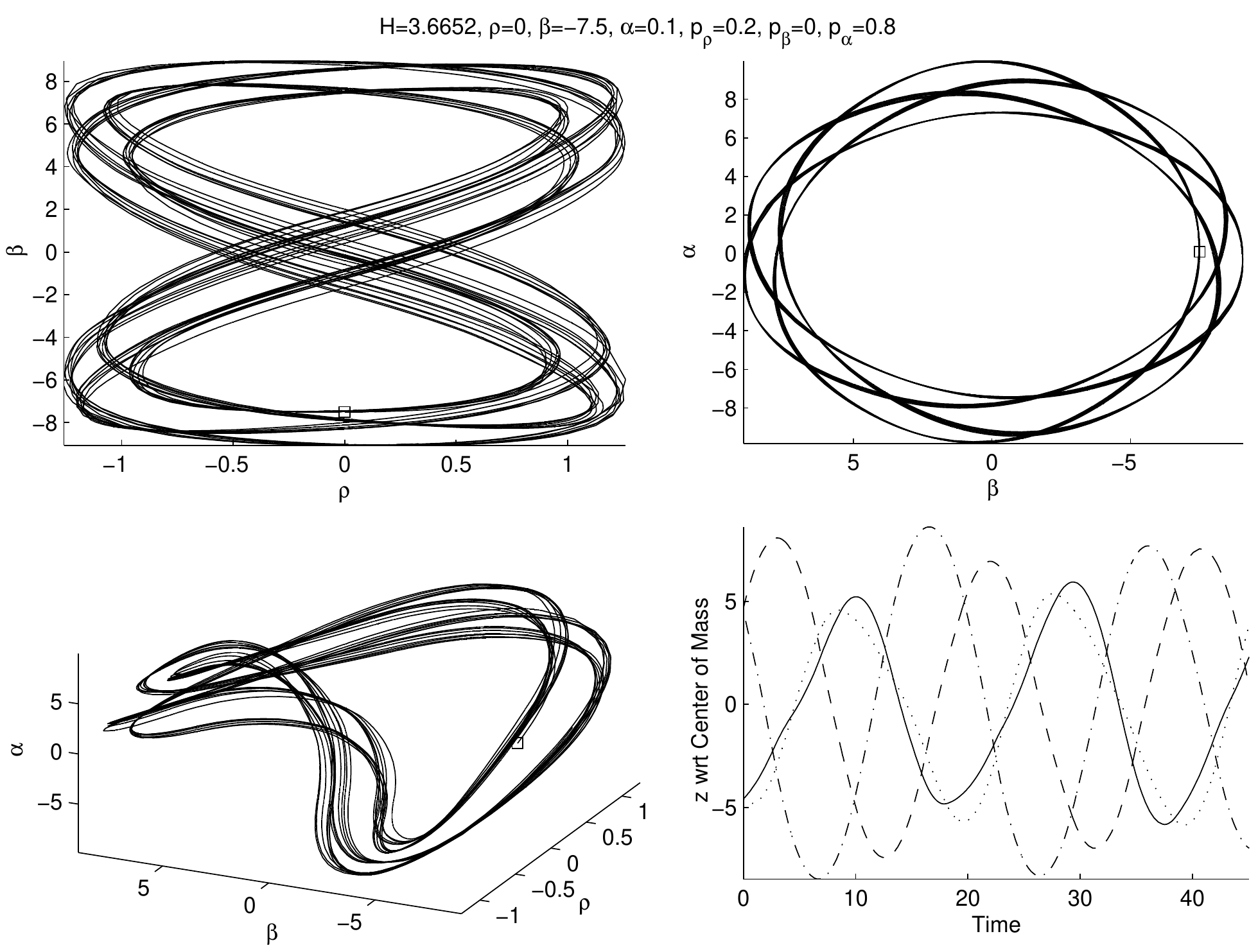}
\par\end{centering}

\begin{centering}
\includegraphics[width=1\linewidth]{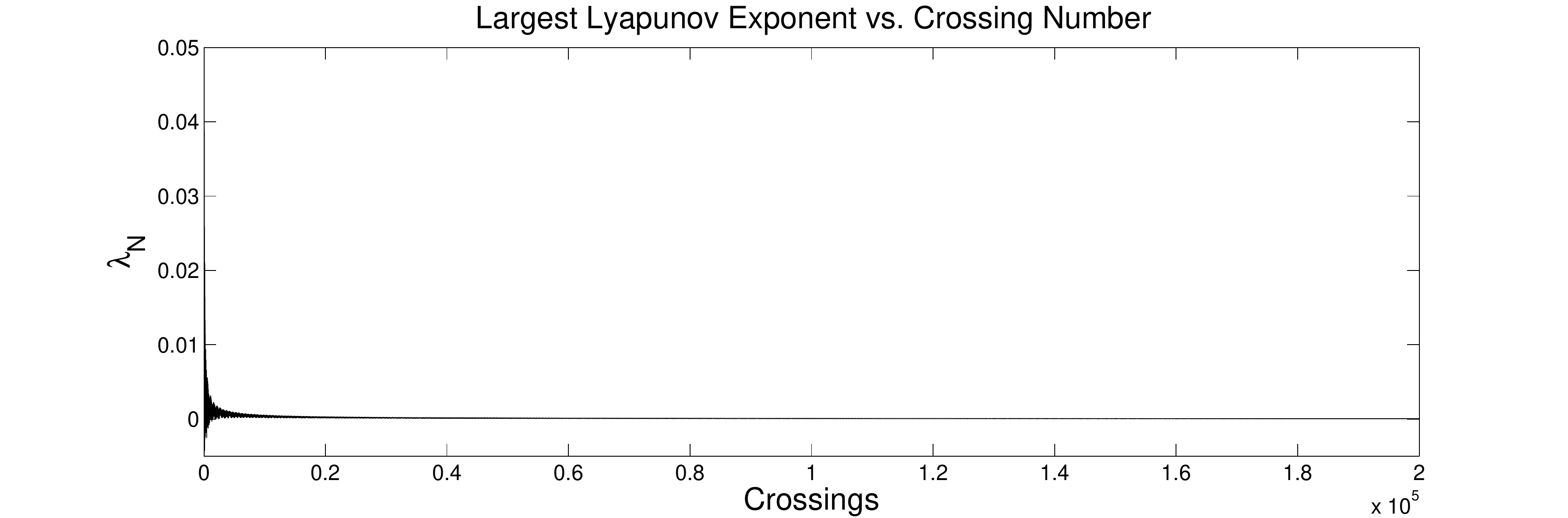}
\par\end{centering}

\caption{Another example of a periodic three-dimensional pa\textcolor{black}{th.
The Lyapunov graph for this trajectory is at the bottom, and its Lyapunov
exponent was calculated to be $3.323\times10^{-5}.$}}

\label{fig:ann_pretz_2} 
\end{figure}

To give an even better idea of the types of paths that can be obtained
with compatible periods, another example is shown in Figure \ref{fig:ann_pretz_2}.
\ The symbol sequence for this trajectory is similar to the other
example in that it contains mostly sets of $CB^{2}$\ and $CB^{6}$;
however $CB^{12}$\
also shows up occasionally. 

Suffice it to say that many interesting trajectories can be obtained,
although most are generalizations of the three-body case, as seen
in the previous section, or are simply chaotic. \ As expected we
find that the latter case produces the same type of dense orbits in
three-dimensions as its 3-body counterpart did in two.

\subsection{Trajectory Plots of Special-Cases}

We will now focus our attention on the various special cases that
we can attain by starting particles very close to one another. \ For
example, we can put two sets of two particles together, and we expect
those two to repeatedly cross one another, while on average the two
pairs behave as a equal-mass two body system. \ Similarly we could
put three particles very close together and one separate.

From this and the fact that we have equal masses we will consider
six initial configurations:\ 1+1+1+1, 2+2, 3+1, 2+1+1, 1+2+1 and
1+1+2, where ``a+b+c''\ notation indicates the relative size of
the initial $z$\ values, with largest at the left and smallest at
the right; a number larger than unity indicates that the difference
between the $z$-values of these particles is small relative to all
other spacings. For example, ``2+1+1'' denotes that we have a pair
of particles starting close together (relative to the other spacings)
with the greatest $z$\ values, and two independent (far apart) particles
with the lower two $z$\ values. \ Strictly speaking the 1+1+2 and
2+1+1 cases are symmetric (equal masses) and indeed they produce similar
plots as we will see.

For each of these possible cases, we construct plots of \ the box-particle
in $(\rho,\beta,\alpha)$ space and the associated plot of particle
positions on the line. \ The easiest way to find suitable initial
conditions is to use (\ref{changevar1}) to choose $\rho,\beta$ and
$\alpha$ so that we get suitable particle starting positions, and
set the momenta to zero. \
Thus we use the fixed-momenta conditions and allow $H$ to vary. \ Only
one angle of the $(\rho,\beta,\alpha)$ plots is shown, though the
captions describe them in more detail.

\begin{figure}[tbph]
\begin{centering}
\includegraphics[width=1\linewidth]{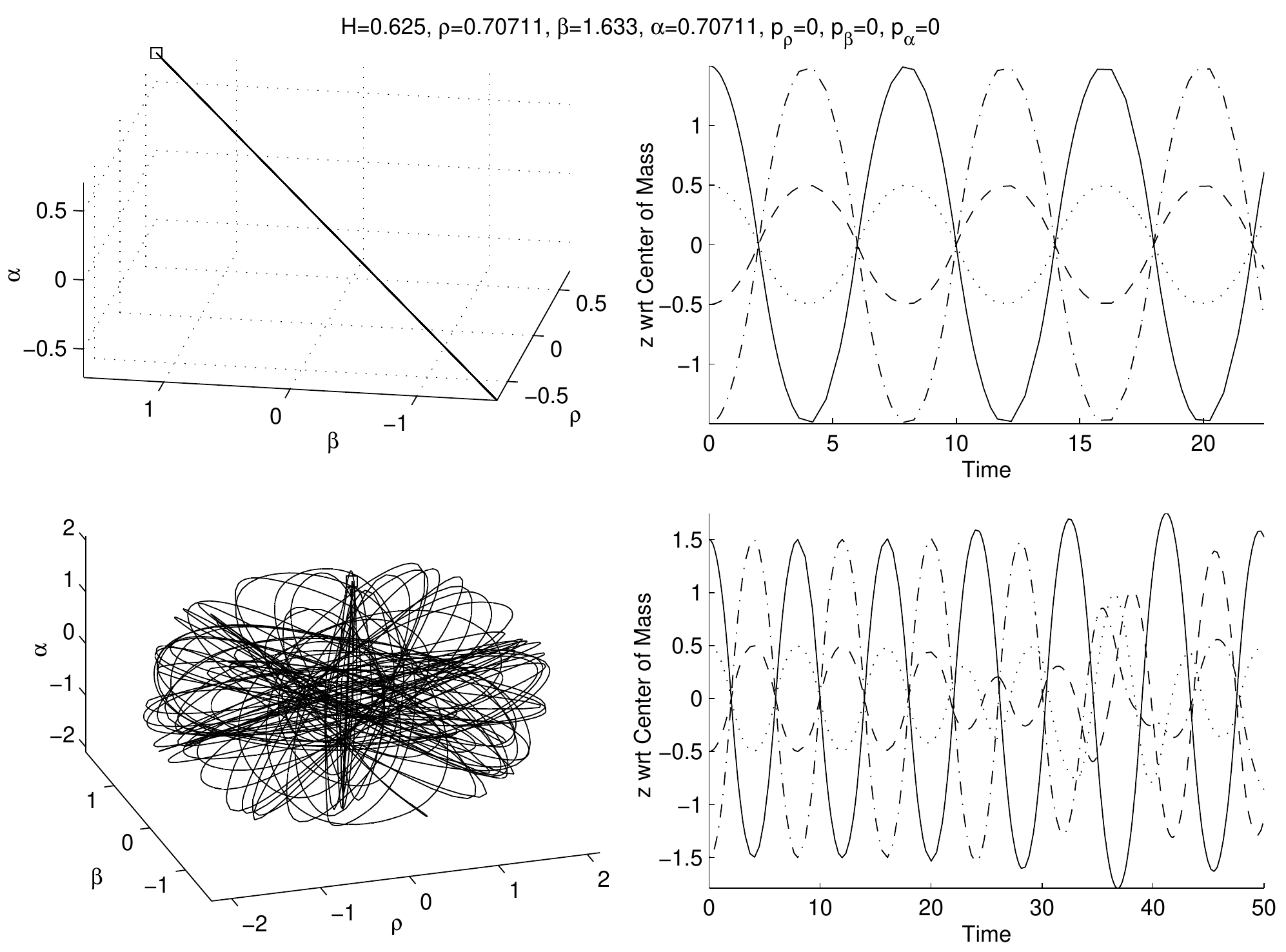}
\par\end{centering}

\begin{centering}
\includegraphics[width=1\linewidth]{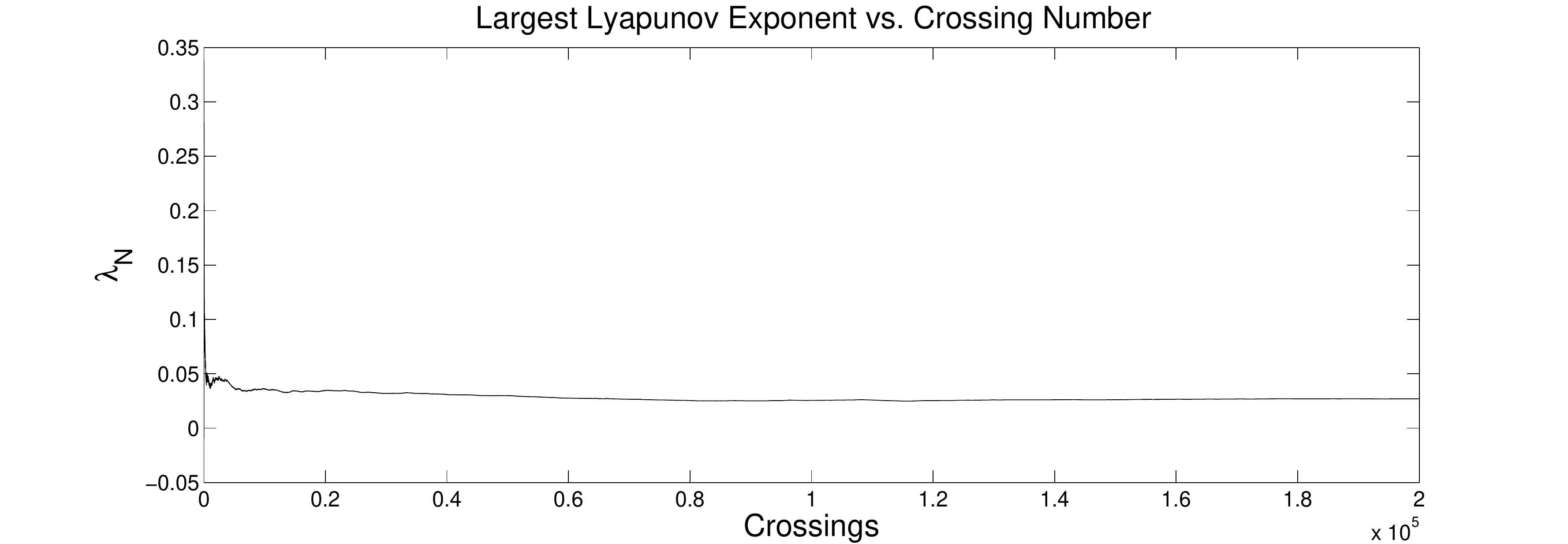}
\par\end{centering}

\caption{The 1+1+1+1 case. \ The $(\rho,\beta,\alpha)$ plot is a single line
diagonal through the three axes, as shown here. \ The bottom two
images show what happens if we change the initial conditions very
slightly (here we set $\rho^{\prime}=\rho+0.0001$). \ A\textcolor{black}{s
can be seen on the $(z,t)$ plot, the system quickly degenerates into
chaos. The Lyapunov graph for the perturbed trajectory, whose Lyapunov
exponent was calculated to be$2.697\times10^{-2}$ appears at the
bottom.}}

\label{fig:1_1_1_1} 
\end{figure}

Figure \ref{fig:1_1_1_1} shows the case where the four particles
begin spaced at equal unit spacing with no momentum. \ Predictably
they are all drawn together and cross at the same point, continuing
this pattern indefinitely. \ Since all four particles always `collide'
at the same time step, the symbol sequence in terms of $A$, $B$,
and $C$\ is undefined. Instead the motion reduces to that of a single
body of mass $4m.$\ \ The box particle travels back and forth along
a line in $(\rho,\beta,\alpha)$\
space, with all particles simultaneously crossing at the origin. This
motion is unstable: a slight change in any of the initial conditions
(either via a small displacement or small momentum) throws the system
into chaos, as shown in the bottom two images of Figure \ref{fig:1_1_1_1}.

\begin{figure}[tbph]
\begin{centering}
\includegraphics[width=1\linewidth]{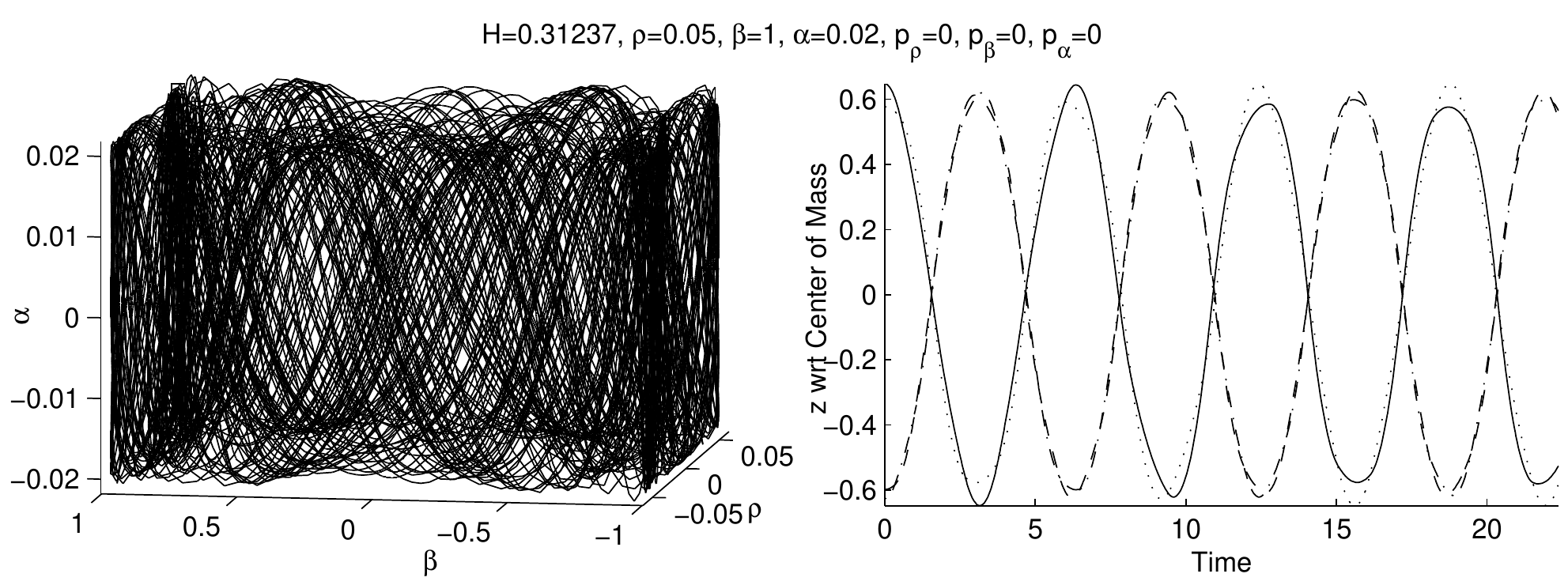}
\par\end{centering}

\caption{The 2+2 case. \ The $(\rho,\beta,\alpha)$ plot is a very dense pretzel
if proj\textcolor{black}{ected onto two of the planes, and an almost
fully filled square in the third. The Lyapunov exponent for this trajectory
was calculated to be $1.307\times10^{-2}.$}}

\label{fig:2_2} 
\end{figure}

In Figure \ref{fig:2_2}, we see the case with two pairs of particles,
denoted as 2+2. Each of the two tightly bound states undergo mildly
irregular motion (where each particle randomly attains the maximal
separation from the origin), but that the orbit of each bound pair
about the other is quite regular, with fairly constant frequency and
amplitude. \
Numerically we estimate the frequency of the oscillation of the pairs
at $0.16$\ cycles per time step, with an amplitude of $1.2$. \ The
motion is a string of $A$'s, $B$'s, and $C$'s without any discernible
pattern, the only noticeable feature being that the $B$'s always
occur in pairs ($B^{2}$). 

\begin{figure}[tbph]
\begin{centering}
\includegraphics[width=1\linewidth]{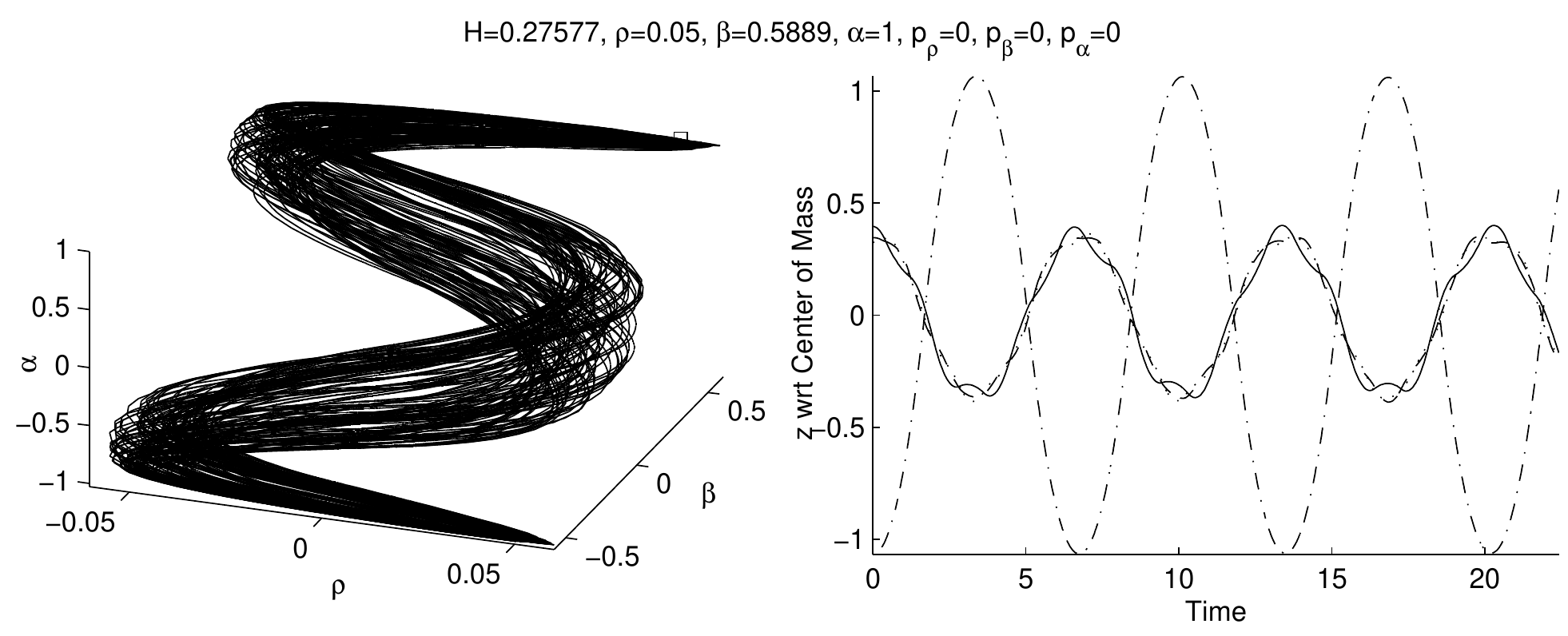}
\par\end{centering}

\begin{centering}
\includegraphics[bb=100bp 15bp 840bp 410bp,clip,width=0.5\linewidth,height=0.4\linewidth]{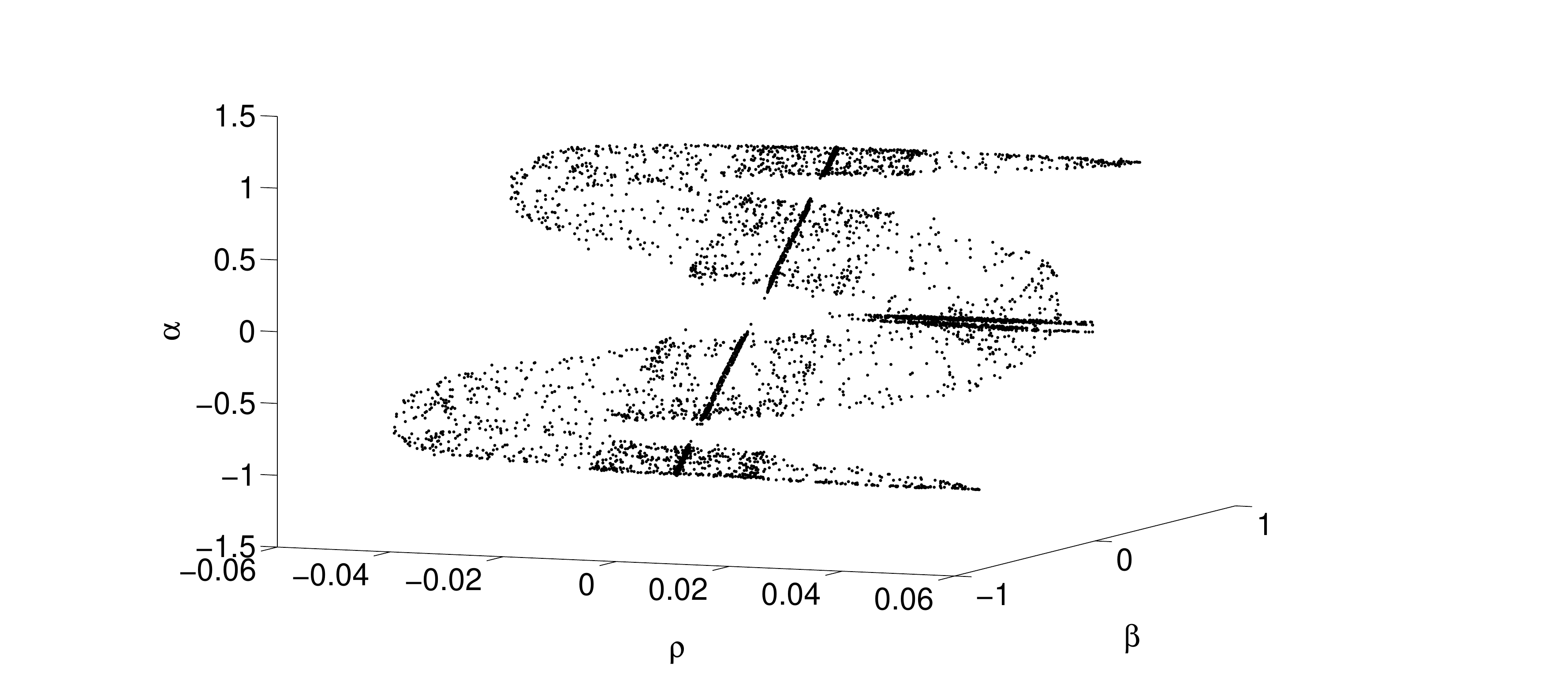}
\par\end{centering}

\begin{centering}
\includegraphics[width=1\linewidth]{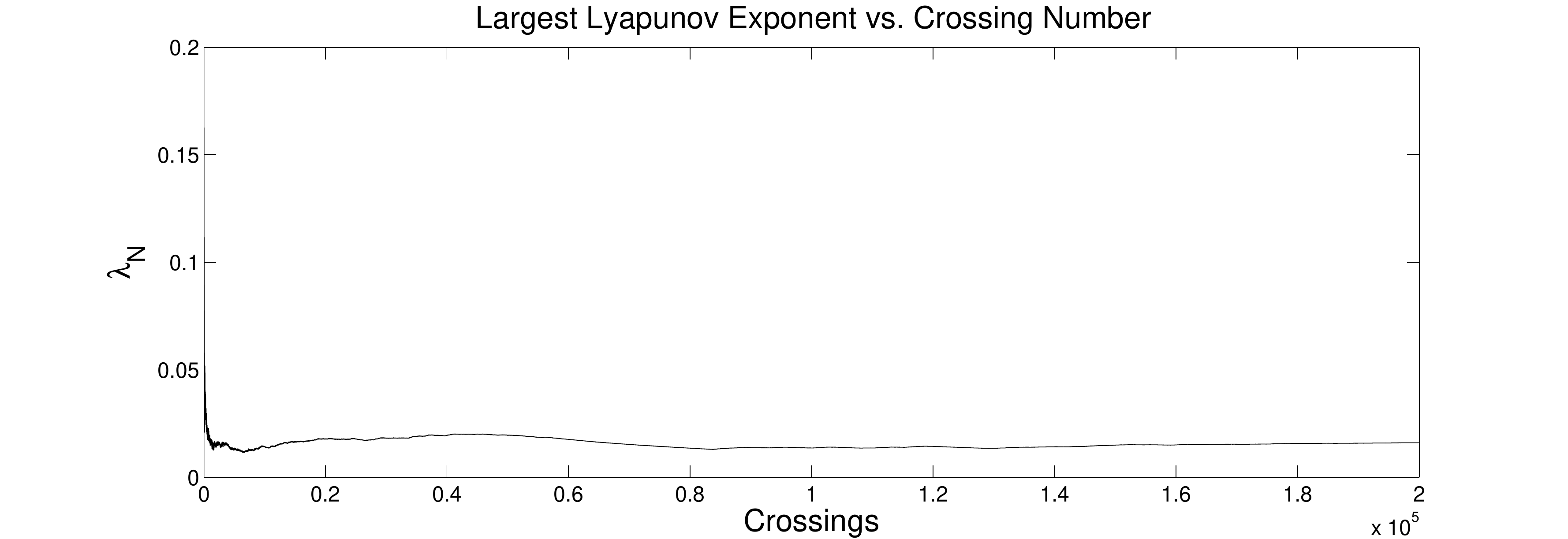}
\par\end{centering}

\caption{The 3+1 case. \ The $(\rho,\beta,\alpha)$ plot is an almost two-dimensional
pretzel rotated sidewa\textcolor{black}{ys. The figure below it was
obtained using the collision to collision mapping solution. Once again,
the dimensions and banding structure are common features of the two
plots, though these features are only faintly visible in the upper
right graph. The bottom figure is the Lyapunov graph for this trajectory,
whose Lyapunov exponent was found to be $1.609\times10^{-2}.$}}

\label{fig:1_3} 
\end{figure}

Figure \ref{fig:1_3}\ shows the \,3+1 case with three particles
closely bound together. \ These undergo a mildly chaotic orbit that
itself regularly oscillates with an amplitude of about $1.4$\ and
a frequency of $0.15$\ cycles per time step about the 4th particle
in a loosely bound state. \textit{\ }The symbol sequence is mostly
$B$'s, with $A$\ or $C$\
occasionally appearing. \ Note that while we still do not see any
$B^{1\text{'}}$s here, we do see some odd numbered groups of $B$'s.

\begin{figure}[tbph]
\begin{centering}
\includegraphics[width=1\linewidth]{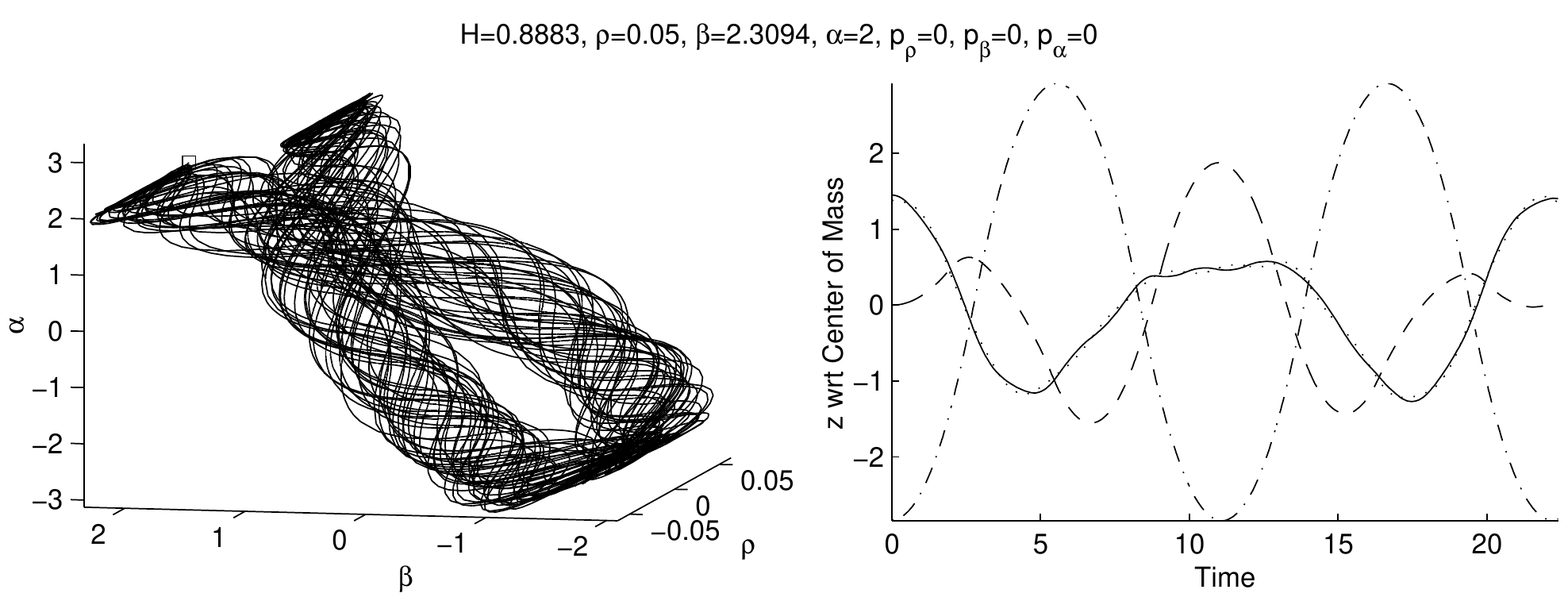}
\par\end{centering}

\caption{The 2+1+1 case. \ The $(\rho,\beta,\alpha)$ plot is a ``fish''
type shape in one plane, and a mess in the others (as seen in the
previo\textcolor{black}{us sections). Similar ``molecular'' structures
have been noted previously in the relativistic \cite{Burnell,justinrobb}
and non-relativistic \cite{Rouet} 3-body cases. The Lyapunov exponent
for this trajectory was calculated to be $2.336\times10^{-2}.$}}

\label{fig:2_1_1} 
\end{figure}

\begin{figure}[tbph]
\begin{centering}
\includegraphics[width=1\linewidth]{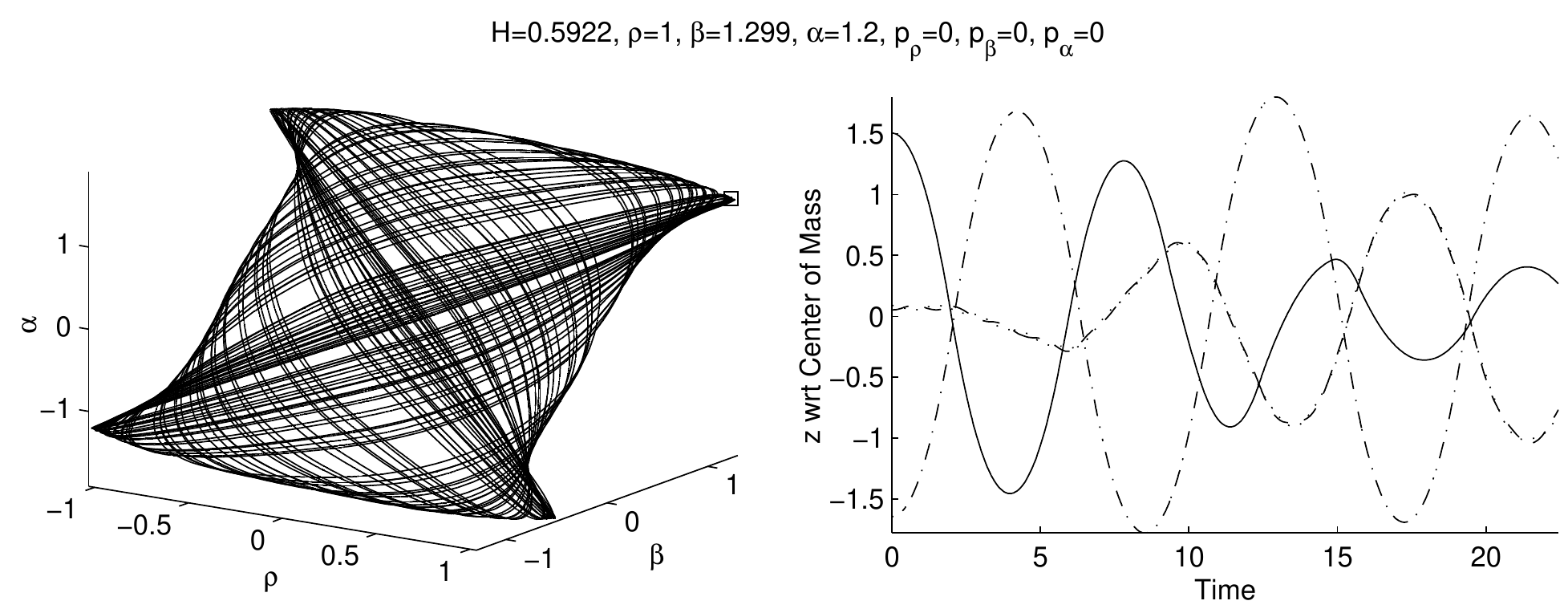}
\par\end{centering}

\caption{The 1+2+1 case. \ The $(\rho,\beta,\alpha)$ plot is an alm\textcolor{black}{ost
two-dimensional pretzel rotated on two axes and skewed on one. The
Lyapunov exponent for this trajectory was calculated to be $1.133\times10^{-2}.$}}

\label{fig:1_2_1} 
\end{figure}

\begin{figure}[tbph]
\begin{centering}
\includegraphics[width=1\linewidth]{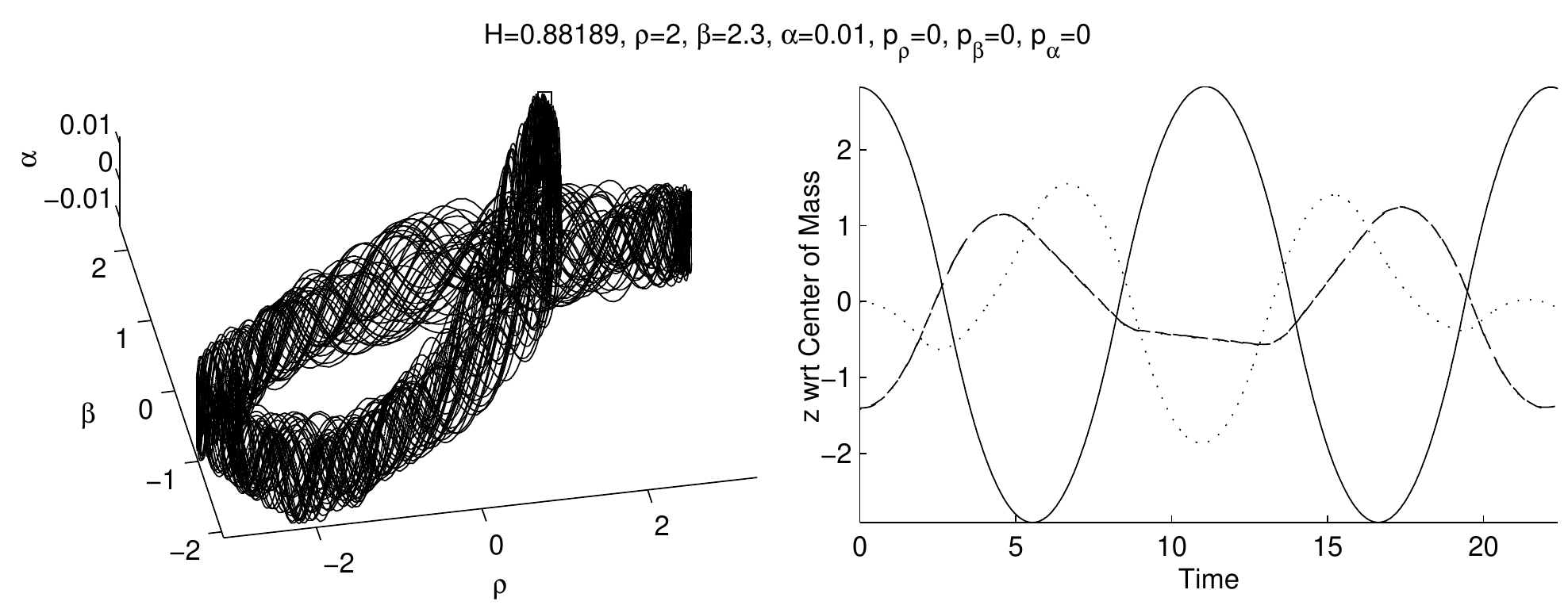}
\par\end{centering}

\caption{The 1+1+2 case. \ Th\textcolor{black}{e $(\rho,\beta,\alpha)$ plot
is a very nice ``fish'' shape in one plane. The Lyapunov exponent
for this trajectory was calculated to be $2.842\times10^{-2}.$}}

\label{fig:1_1_2} 
\end{figure}

Figures \ref{fig:2_1_1}-\ref{fig:1_1_2} show the cases where only
one pair of particles is closely bound, namely 2+1+1, 1+2+1 and 1+1+2
respectively. Fig. \ref{fig:2_1_1} generates a thick `fish' pattern:
the tightly bound state of two particles executes a $\overline{B^{4}A}$\ motion
with respect to the other two. \ However the full 4-body sequence
does not exhibit any clear pattern.

Fig. \ref{fig:1_2_1} illustrates a similar kind of motion, in which
the tightly bound state undergoes a $\overline{B^{10}A}$\ motion
with respect to the other two; the time period shown on the graph
is too short to see this explicitly in the figure. \ Again the 4-body
crossing sequence has no discernible pattern.

In Fig. \ref{fig:1_1_2} we see another similar trajectory, but with
subtly distinct features. \ The two particles are not as tightly
bound as in the preceding cases, but (over the time scales we observed)
execute a highly regular oscillatory interaction with the other two
particles. \ The four particles repeatedly come very close together
before executing near parabolic motion about the centre of mass.\ \ However
after about 80 time steps the trajectory degenerates into chaos. \ While
the paired particles remain bound, they eventually fall out of the
parabolic motion and begin to move chaotically with respect to the
other particles. \ No pattern is obvious in the symbol sequence.

\subsection{Poincare Plots}

We will now examine the Poincare plots of these trajectories in the
Newtonian system. \ Unfortunately, this turns out to be much more
difficult in the four-body case than it was in the three-body. \ The
reason is that in the three-body system, the Poincare plots were a
collection of points in two-dimensions \cite{Burnell,justinrobb},
while in the four-body system they are a collection of points in \textit{three}
dimensions. \ The latter collection is much more difficult to visualize
since it is impossible for the human eye to determine the depth of
a single point on these three-dimensional plots with sufficient precision.
\ Furthermore, the outer layers of points tend to obscure inner ones
exists. \ Thus we will examine the Poincare plots of several independent
three-dimensional trajectories, and then look at ways of visualizing
the ``total'' Poincare plot, with a broad range of initial conditions
combined.

Poincare plots are plots of sections of phase space, affording a comparison
of classes of trajectories over a broad range of initial conditions.
We shall set $H=1$ throughout. In the equal mass case all of the
bisectors of the 3-simplex represent equivalent crossings of particles.
\ Thus we can plot the crossing of any two particles (equivalent
to the $(\rho,\beta,\alpha)$ particle crossing one of the bisectors
in the three-dimensional potential) all on the same graph. 

In the three-body case the Poincare plot was that of the radial momentum
of the simplex particle plotted against its angular momentum every
time the particle crossed a simplex bisector \cite{LMiller,Burnell,justinrobb}.
The most natural extension of the construction of these plots to three
dimensions is to use spherical coordinates, and plot the radial momentum,
denoted here as $p_{R},$ against the two angular momenta squared:
$p_{\phi}^{2}$ and $p_{\theta}^{2}$. \ More concisely, we define
$R$ to be the distance from the origin to our point of crossing in
$(\rho,\beta,\alpha)$ space, $\phi$ to be the \textit{azimuthal}
angle in the $(\rho,\beta)$ plane (with $0\leq\phi\leq2\pi$), and
$\theta$ to be the \textit{polar} angle from the $\alpha$ axis (with
$0\leq\theta\leq\pi$). \ From this, we denote the associated momenta
of $R,\phi$ and $\theta$ as $p_{R},p_{\phi}$ and $p_{\theta}$
respectively.\ \ 

Specifically, from simple geometry relating our spherical coordinates
$(R,\phi,\theta)$ to our Cartesian coordinates $(\rho,\beta,\alpha)$,
we have that
\begin{eqnarray}
\sin\phi & = & \frac{\beta}{\sqrt{\rho^{2}+\beta^{2}}},\quad\quad\quad\;\cos\phi=\frac{\rho}{\sqrt{\rho^{2}+\beta^{2}}}\label{spheretrig}\\
\sin\theta & = & \frac{\sqrt{\rho^{2}+\beta^{2}}}{\sqrt{\rho^{2}+\beta^{2}+\alpha^{2}}},\quad\cos\theta=\frac{\alpha}{\sqrt{\rho^{2}+\beta^{2}+\alpha^{2}}}\notag
\end{eqnarray}
and the unit vectors for these spherical coordinates are:

\begin{equation}
\hat{R}=\left[\begin{array}{c}
\cos\phi\sin\theta\\
\sin\phi\sin\theta\\
\cos\theta
\end{array}\right],\quad\hat{\phi}=\left[\begin{array}{c}
-\sin\phi\\
\cos\phi\\
0
\end{array}\right],\quad\hat{\theta}=\left[\begin{array}{c}
\cos\phi\cos\theta\\
\sin\phi\cos\theta\\
-\sin\theta
\end{array}\right]\label{sphereuv}
\end{equation}

The desired momenta are:

\begin{equation}
p_{R}=\hat{R}\cdot\vec{p},\qquad p_{\phi}=\hat{\phi}\cdot\vec{p},\qquad p_{\theta}=\hat{\theta}\cdot\vec{p}\label{spheremom1}
\end{equation}
where $\vec{p}\equiv(p_{\rho},p_{\beta},p_{\alpha})$ is the momentum
vector in $(\rho,\beta,\alpha)$ space. \ Then by substituting (\ref{spheretrig})
and (\ref{sphereuv}) into (\ref{spheremom1}), we can find an expression
for the momenta in terms of $\rho,\beta,\alpha,p_{\rho},p_{\beta}$
and $p_{\alpha}$:
\begin{eqnarray}
p_{R} & = & \frac{p_{\rho}\rho+p_{\beta}\beta+p_{\alpha}\alpha}{\sqrt{\rho^{2}+\beta^{2}+\alpha^{2}}}\notag\\
p_{\phi} & = & \frac{-p_{\rho}\beta+p_{\beta}\rho}{\sqrt{\rho^{2}+\beta^{2}}}\label{spheremom2}\\
p_{\theta} & = & \frac{p_{\rho}\rho\alpha+p_{\beta}\beta\alpha-p_{\alpha}(\rho^{2}+\beta^{2})}{\sqrt{(\rho^{2}+\beta^{2}+\alpha^{2})(\rho^{2}+\beta^{2})}}\notag
\end{eqnarray}

Thus we can use (\ref{spheremom2}) to plot $p_{R}$ , $p_{\phi}^{2}$
and $p_{\theta}^{2}$ whenever two of the four particles cross one
another.

\subsubsection{Single Trajectory Poincare Plots}

As we will discover further on, it becomes very difficult to visualize
a complete Poincare plot in three dimensions. We therefore will build
up to this case by first examining individual Poincare plots generated
from single t\textcolor{black}{rajectories. Note that all of the following
Poincare plots were produced using both methods and found to be exactly
the same, checking the validity of the following results.}

To keep things simple, we will first look at the case where our trajectory
is in two dimensions. \ This should result in a two dimensional Poincare
plot as well, as in the three-body case. \ Since $p_{\alpha}$ is
varied to keep the initial conditions consistent, we will arbitrarily
choose $\beta$ and $p_{\beta}$ as the variables that we set to zero,
thus making it into a two dimensional trajectory as seen in previous
sections.

\begin{figure}[tbph]
\begin{centering}
\includegraphics[width=1\linewidth]{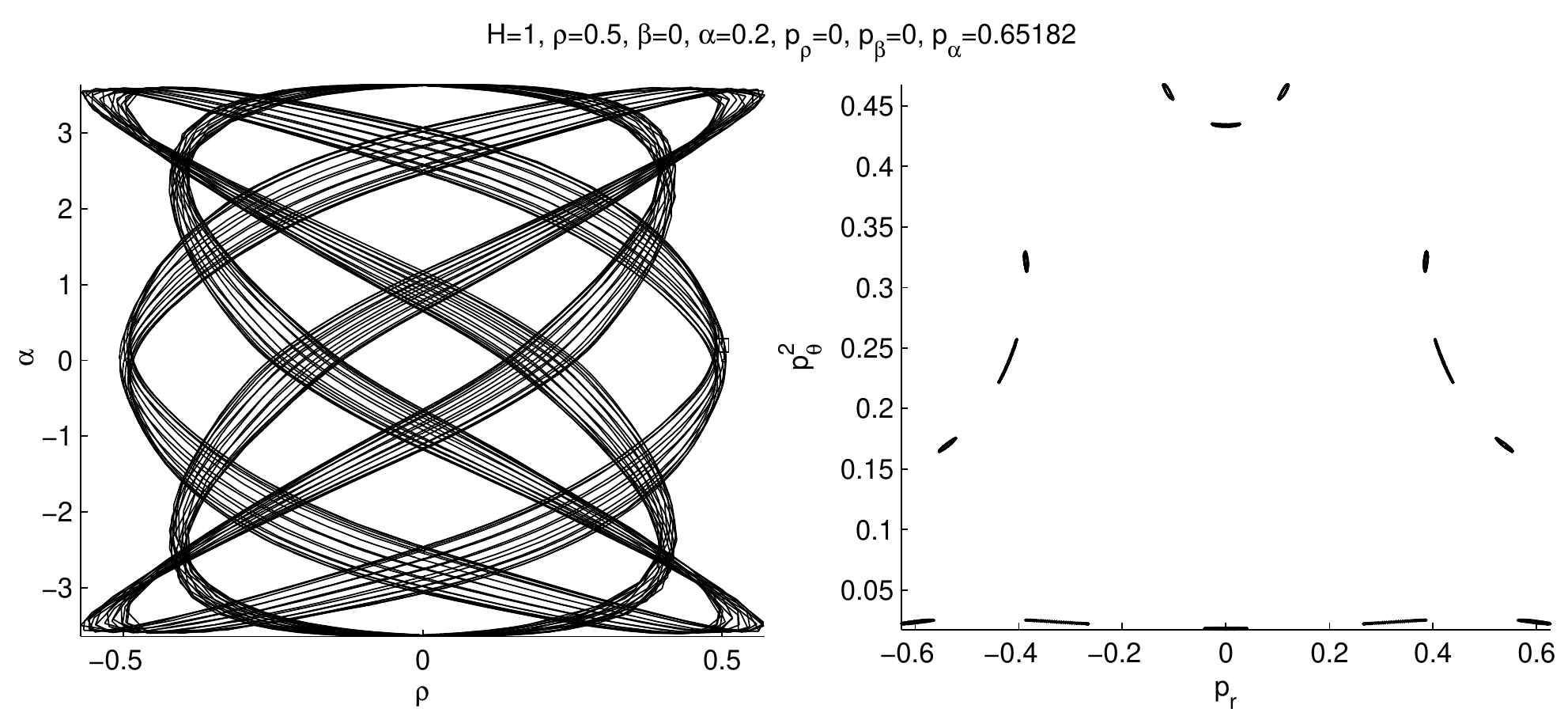}
\par\end{centering}

\caption{The trajectory and Poincare plots for a two-dimensional pretzel. \
Note that the trajectory is shown for five hundred time steps, while
the Poincare section is generated from five \textit{thousand} time
steps. \ The latter is required to generate enough points from a
single trajectory to show a meaningful plot.}

\label{fig:2d_pretz_poinc} 
\end{figure}

Our intuition is indeed correct, as can be seen in Figure \ref{fig:2d_pretz_poinc}.
\ There a two-dimensional pretzel shape produces a two-dimensional
Poincare plot that coincides with those seen in the three-body case,
consistent with the quasiperiodic motion of this trajectory.\ As
expected, the value of the $p_{\phi}$\ (the momenta of the azimuthal
angle) remains zero throughout. \ This can also be seen from (\ref{spheremom2}),
with $\beta=0$ and $p_{\beta}=0$. More importantly, this confirms
that the Poincare plots of the four-body system will be generalizations
of the three-body system, as was seen with the trajectories, validating
our approach for generating these plots. 

\begin{figure}[tbph]
\begin{centering}
\includegraphics[width=1\linewidth]{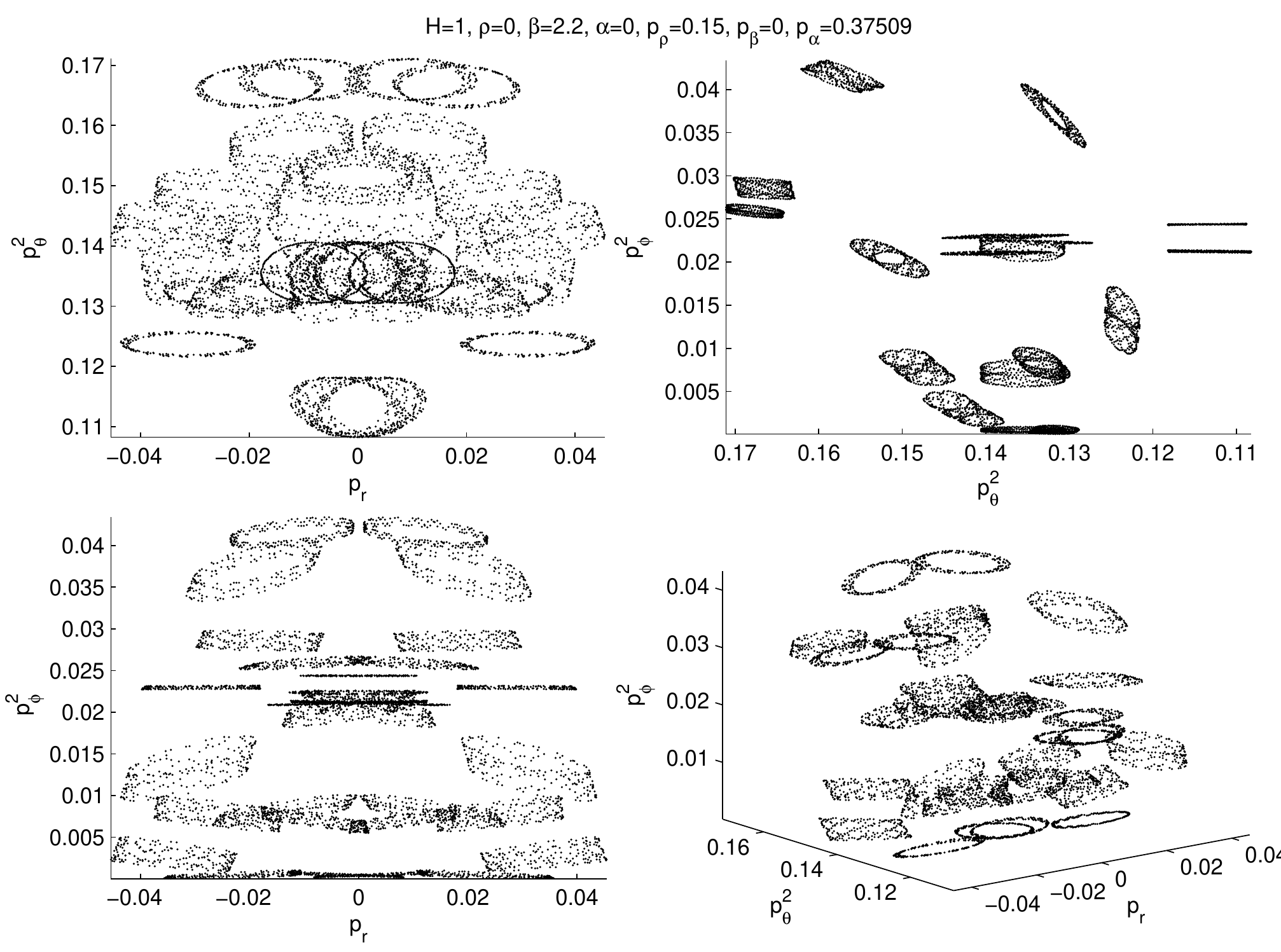}
\par\end{centering}

\caption{The Poincare plot of the trajectory in Figure \ref{fig:ann_pretz},
shown here from all three directions, and from an isometric perspective.
\
The broad circular bands are indicative of the quasiperiodic character
of this trajectory.}

\label{fig:ann_pretz_poinc} 
\end{figure}

Now, we will look at the Poincare plots for some of the ``nicer''
three-dimensional trajectories. \ From Figure \ref{fig:ann_pretz}
we generate a Poincare plot that is given in figure \ref{fig:ann_pretz_poinc},
shown from a few angles.

This plot is extremely interesting in that it does not seem to exhibit
the same patterns as the three-body case. \ Remembering that this
shape is actually a combination of annuli and pretzels, we can begin
to see some patterns. \ First, the elliptical shells (they are actually
extruded ellipses) in this plot are very similar to the ellipses in
the pretzel section of the three-body Poincare plots, although no
geometric pattern is immediately obvious. \ Second, if we compare
the scale on this plot to that of the previous and final Poincare
plots (see next section), we discover that indeed it is only occurring
in a smaller region of the larger plot - potentially the three-dimensional
equivalent of the ``pretzel region'' seen in the three-body case
(albeit more complex, since this trajectory is not exclusively a pretzel).

\begin{figure}[tbph]
\begin{centering}
\includegraphics[width=1\linewidth]{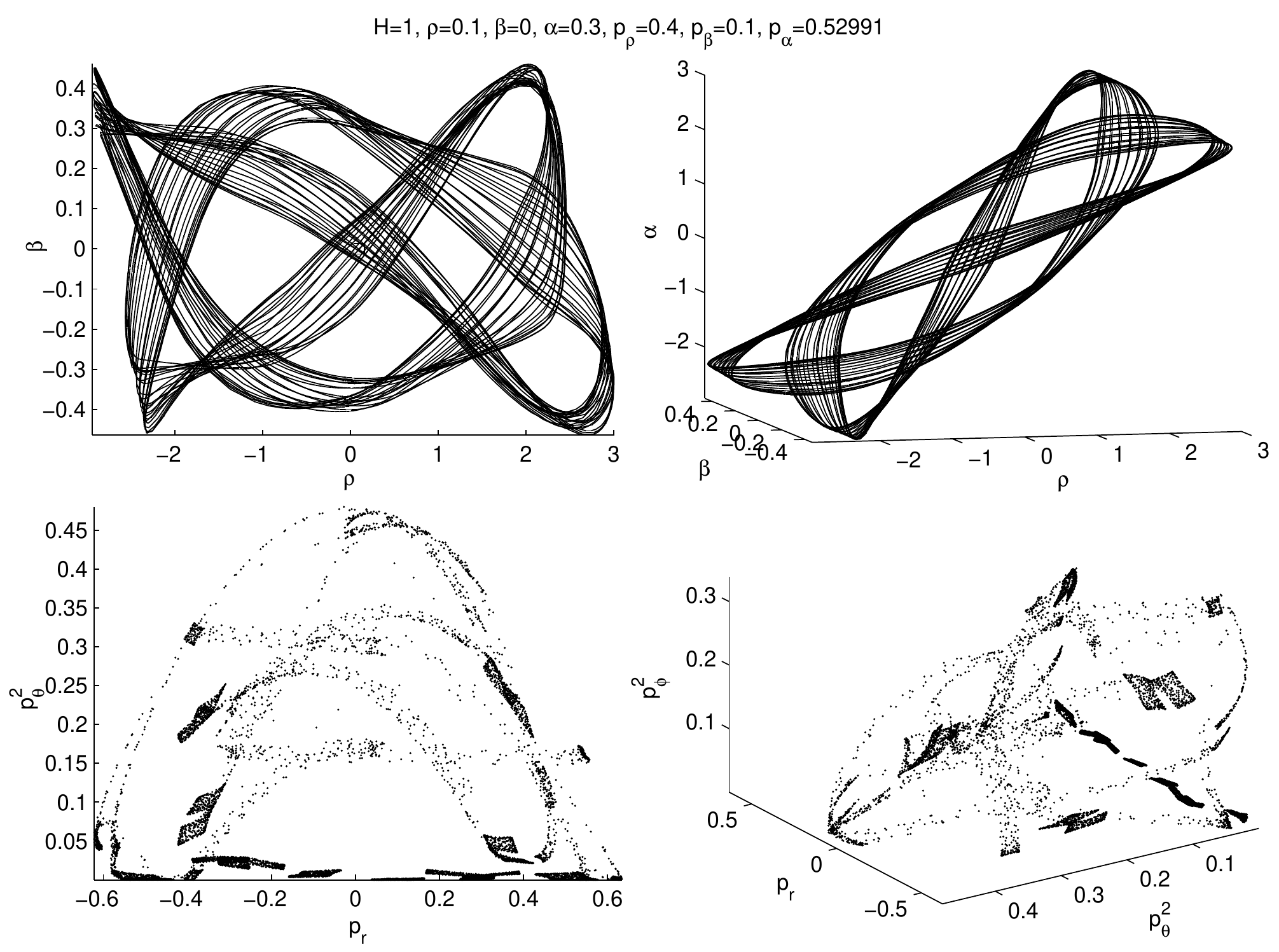}
\par\end{centering}

\caption{The top two images show the trajectory plot, while the bottom two
show the corresponding Poincare plot. \ Although it is difficult
to see on static images, the more loosely dispersed dots form a sort
of ``shell'' over the top of the Poincare plot. \ We can already
see the trouble with visualizing this, even with only a single trajectory
sho\textcolor{black}{wn! The Lyapunov exponent for this trajectory
was calculated to be $4.700\times10^{-4}.$}}

\label{fig:odd_shape} 
\end{figure}

Another example is shown in Figure \ref{fig:odd_shape}. \ We can
see here that because of the skew of the shape in the trajectory plot,
the Poincare plot is not nearly as structured as the previous ones.
\ Still, we can note some pattern, although it is much more difficult
to see in three dimensions. Each trajectory is quasiperiodic, but
spread over a three dimensional space, with the pretzel trajectories
forming a non-overlapping knotted structure. As with the 3-body case,
the Poincare plots consist of circles with some width. These occupy
a three dimensional volume, and the more loosely dispersed dots form
a ``shell'' over the top of the Poincare plot.

Examining other Poincare plots of three-dimensional trajectories,
we reach the limits of this particular approach. \ As we saw earlier,
very few initial conditions actually produce nice three-dimensional
paths, and the ones that do not produce similarly uninformative Poincare
plots. \ It is possible that if run for a significant number of time
steps these plots would materialize into more than just a random collection
of points in space, but the amount of computation required to produce
the hundreds of thousands of time steps that would be required to
check this is daunting, and probably not the best use of resources.
\ Moreover, it becomes increasingly difficult to visualize these
plots as the number of time steps increases. \ 

However, there is some support for the conjecture that more time steps
will reveal better patterns. \ More time steps yield more particle
crossings, and thus more points on the Poincare plot. \ Instead of
extending the length of the trajectory though, we can instead choose
initial conditions such that two particles are extremely close together
and cross frequently.

\begin{figure}[tbph]
\begin{centering}
\includegraphics[width=1\linewidth]{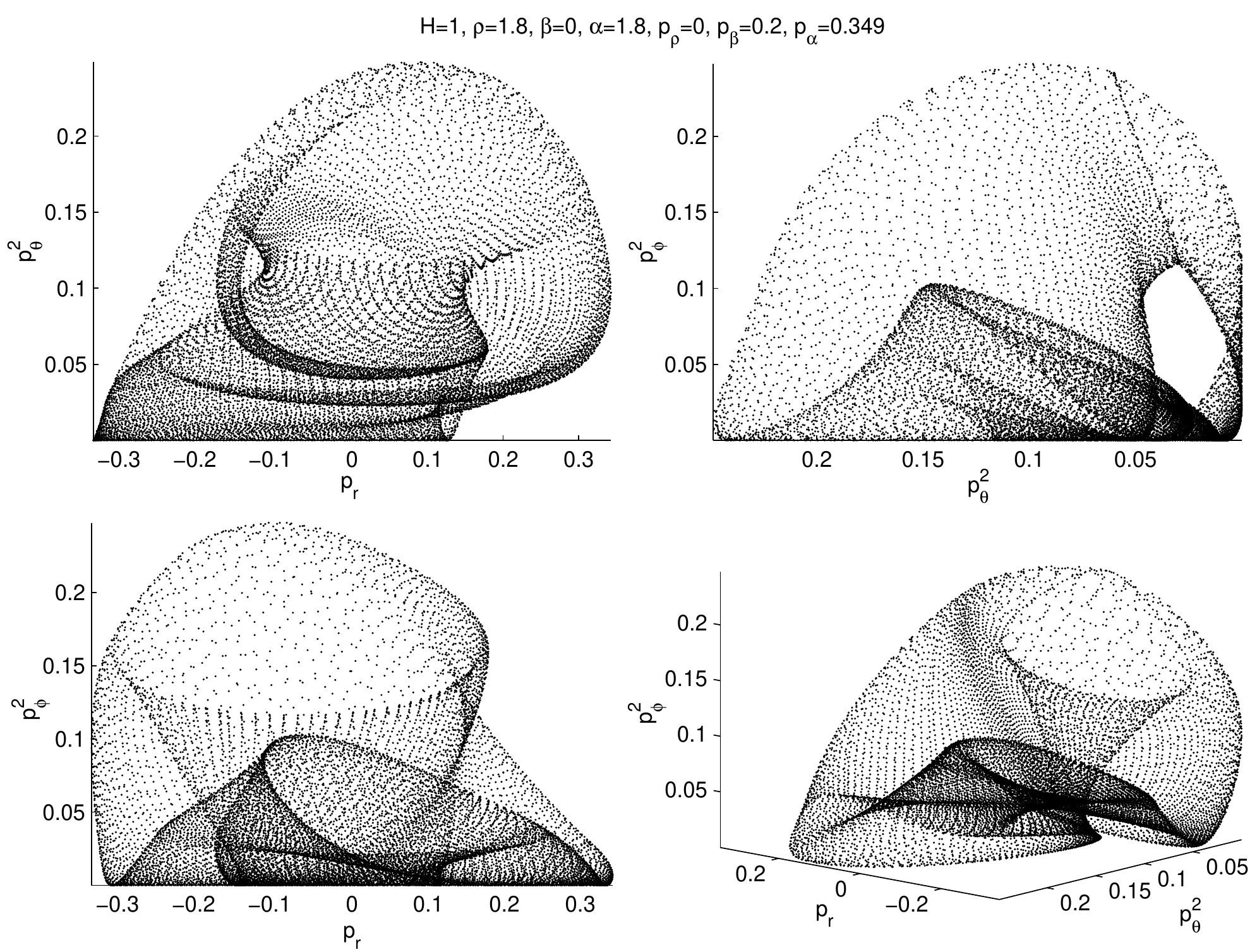}
\par\end{centering}

\caption{The Poincare plot of a highly refined trajectory. \ The box-particle
trajectory is uninteresting:\ it is basically just a two-dimensional
annulus rotated approximately forty-five degrees in two axes. \ Notice
the com\textcolor{black}{plex patterns displayed even in the Poincare
plot for a single trajectory! The Lyapunov exponent for this trajectory
was calculated to be $2.602\times10^{-3}.$}}

\label{fig:nice_poinc} 
\end{figure}

We illustrate this with an example in figure \ref{fig:nice_poinc}.\textit{\ }\ The
motion of the box-particle is characterized by a huge number of $A$'s
with an occasional $B$\ and $C$\ (as we would expect due to the
large number of collisions between the two tightly bound particles).
\textit{\ }This trajectory produces a very nice three dimensional
shape, demonstrating \
that three-dimensional Poincare plots can be considerably more complicated
than their two-dimensional counterparts.

\begin{figure}[tbph]
\begin{centering}
\includegraphics[width=1\linewidth]{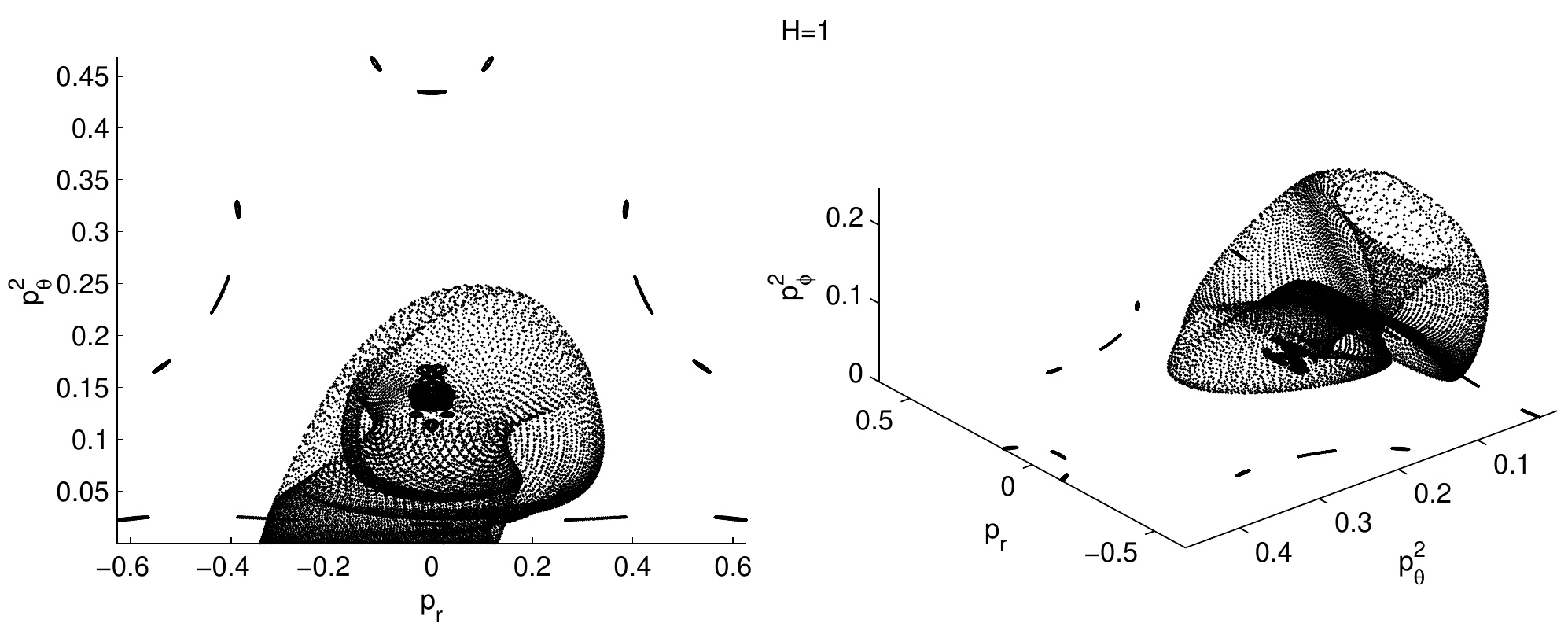}
\par\end{centering}

\caption{The top and isometric views of the combined Poincare plots in this
section. \ Note the relative scale.}

\label{fig:poinc_comb} 
\end{figure}

To conclude this section, we plot a graph (figure \ref{fig:poinc_comb})
that contains all of the previous Poincare plots combined, just to
give an idea of scale.

\subsubsection{3D Poincare plots}

Now we move on to the problem of visualizing the complete Poincare
plot, with a variety of initial conditions included. \ The traditional
approach generally involves choosing a range of initial conditions
that fill in the important regions, and plotting all of the points
on a combined two-dimensional plot. \ This approach has two immediate
problems in three dimensions.

First, the amount of space to fill in three - as opposed to two -
dimensions makes the task of manually choosing initial conditions
formidable. \ There is no reasonable way of ensuring that all of
the possibilities have been covered, especially since there are a
large number of variations and combinations of three-dimensional trajectories.
\ We attempted to address this problem by automating the generation
of data over a specific range of initial conditions. With five independent
variables the number of possible plots is very large, necessitating
a significant reduction in the number of time steps for each initial
condition. \ We generated Poincare data for each trajectory for five
hundred time steps, giving about 1,000 Poincare points per trajectory.
The second problem is that of effectively visualizing all of these
discrete points in three-dimensional space. We approached this problem
by converting the data into volumetric density data, separating out
the space into millions of tiny three-dimensional boxes, assigning
a position corresponding to the location of the box and a value corresponding
to the number of Poincare points that fall inside. \ This yields
a large three-dimensional grid of values, each representing the density
of points in the Poincare plot at that approximate location.

The advantage of this approach is that we can now easily take slices
and contours, colouring them according to the volumetric value at
the corresponding locations. This effectively decomposes the three
dimensional data into a series of two dimensional plots. However there
are disadvantages, the most significant being that we are more limited
in our ability to zoom in to see self-similar patterns and structures
at the resolution the number of chosen grid points. \ Although we
can always use more grid points, we quickly arrive at a situation
in which there are too few points in a small region to satisfactorily
``fill up'' the density data. However we expect that some of the
other important features, like the highly saturated regions of chaos,
will show up well in our new plots. \ Since those regions have a
large number of points, they should get a very high density value
and thus stand out from the surrounding region. \ With enough points
and a small enough grid, it should also still be possible to see the
general shape of the Poincare plot in three dimensions.

We will first examine two special slices:\ that of $p_{\phi}=0$\ (the
``bottom'' slice) and $p_{\theta}=0$\ (the ``side'' slice).

\begin{figure}[tbph]
\begin{centering}
\includegraphics[width=1\linewidth]{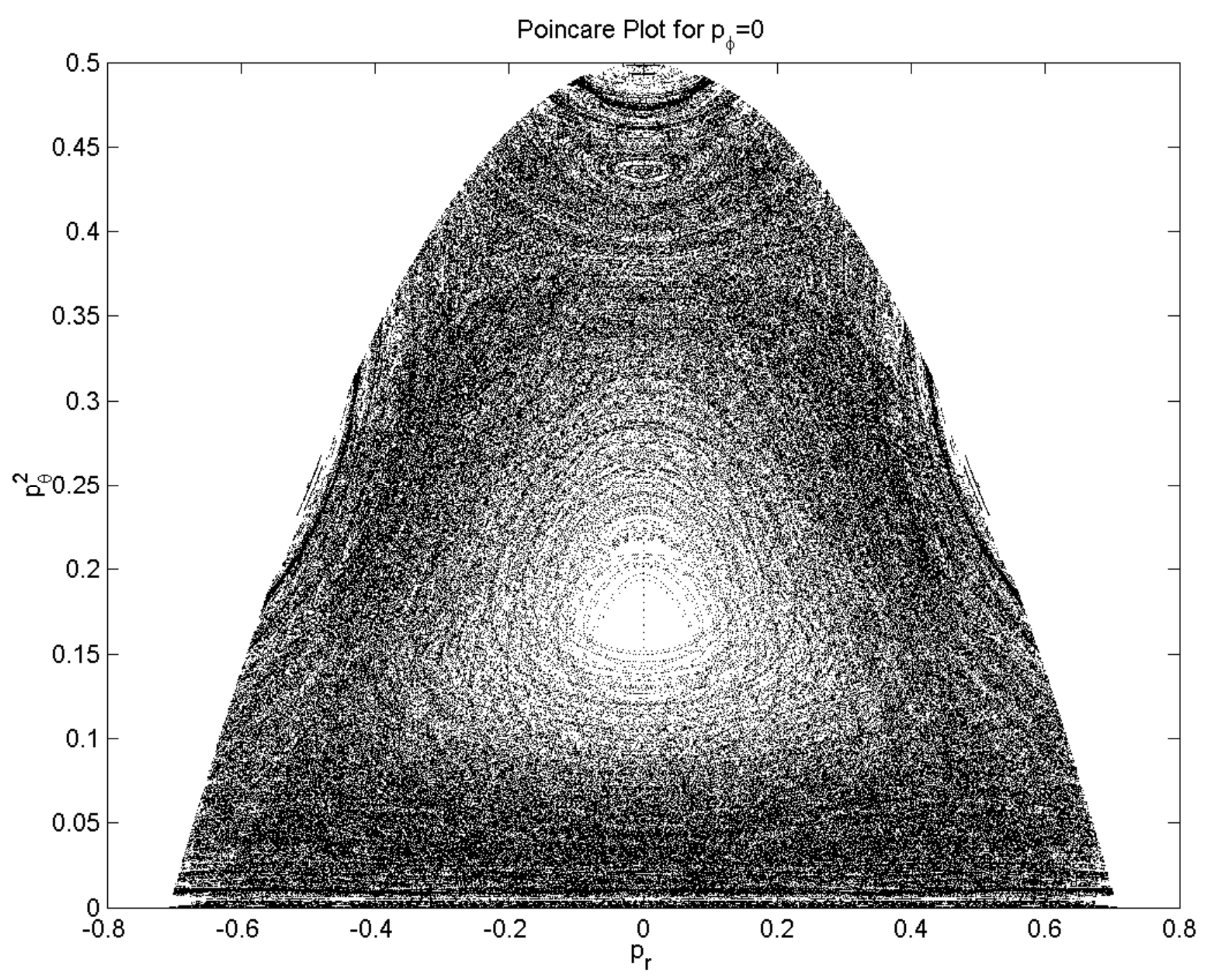}
\par\end{centering}

\caption{The bottom slice of the complete Poincare plot (\symbol{126}400,000
points).}

\label{fig:poinc_tot_bottom} 
\end{figure}

For the bottom slice, we can use the condition $p_{\phi}=0$\ to
select the applicable points. This yields 400,000 points out of the
12 million that we generated. \ As we see in Figure \ref{fig:poinc_tot_bottom},
this two-dimensional slice compares quite well to the 3-body case
\cite{Burnell}. \ The figure suggests the presence of mixed regions
of chaos and integrability, as further implied by figures 20 and 21.

\begin{figure}[tbph]
\begin{centering}
\includegraphics[width=1\linewidth]{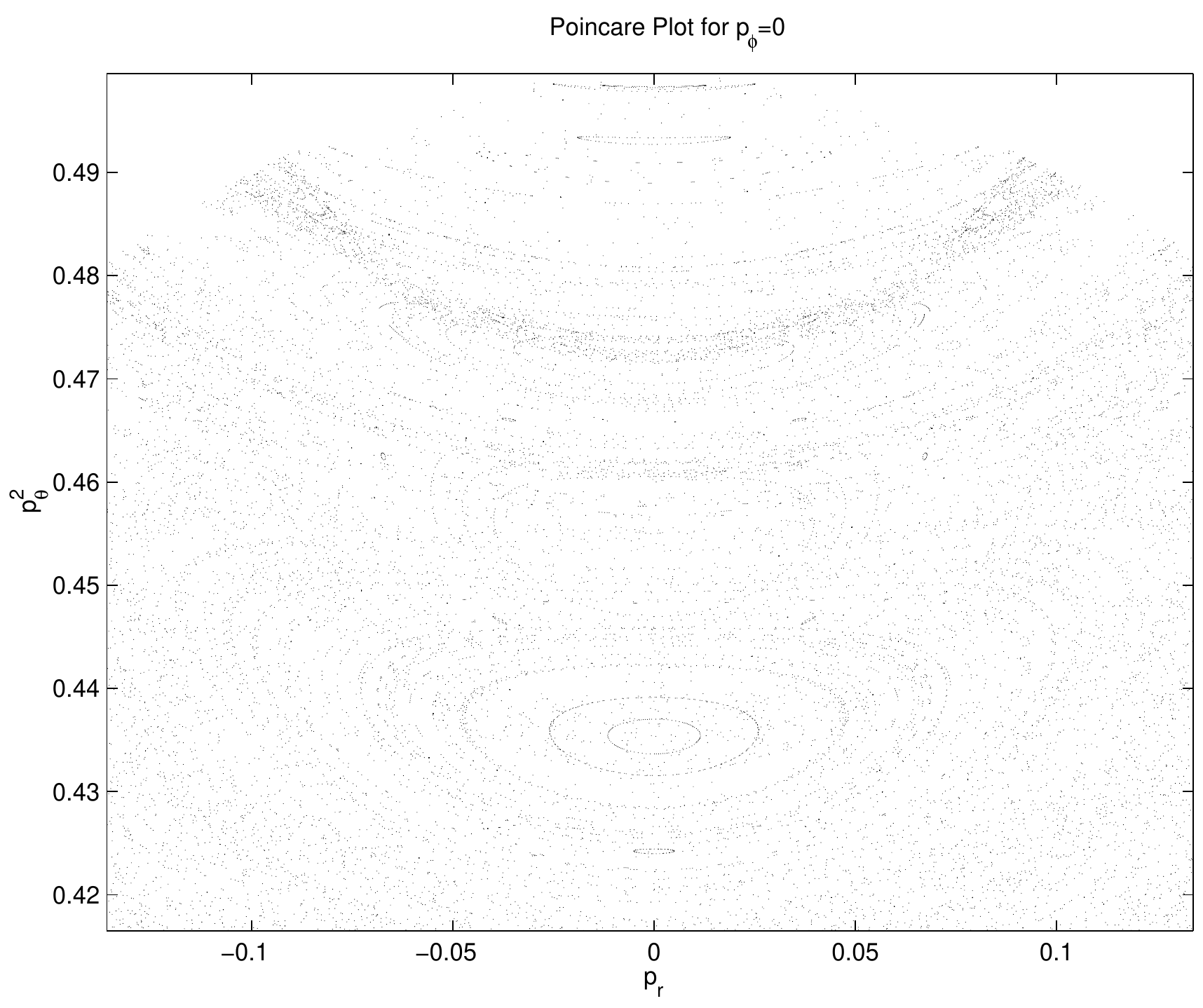}
\par\end{centering}

\caption{A zoomed-in image of the top portion of Figure \ref{fig:poinc_tot_bottom}.
\ Notice the complex fractal-like patterns that show up as we zoom
closer.}

\label{fig:poinc_tot_bottom_zoomed_top} 
\end{figure}

\begin{figure}[tbph]
\begin{centering}
\includegraphics[width=1\linewidth]{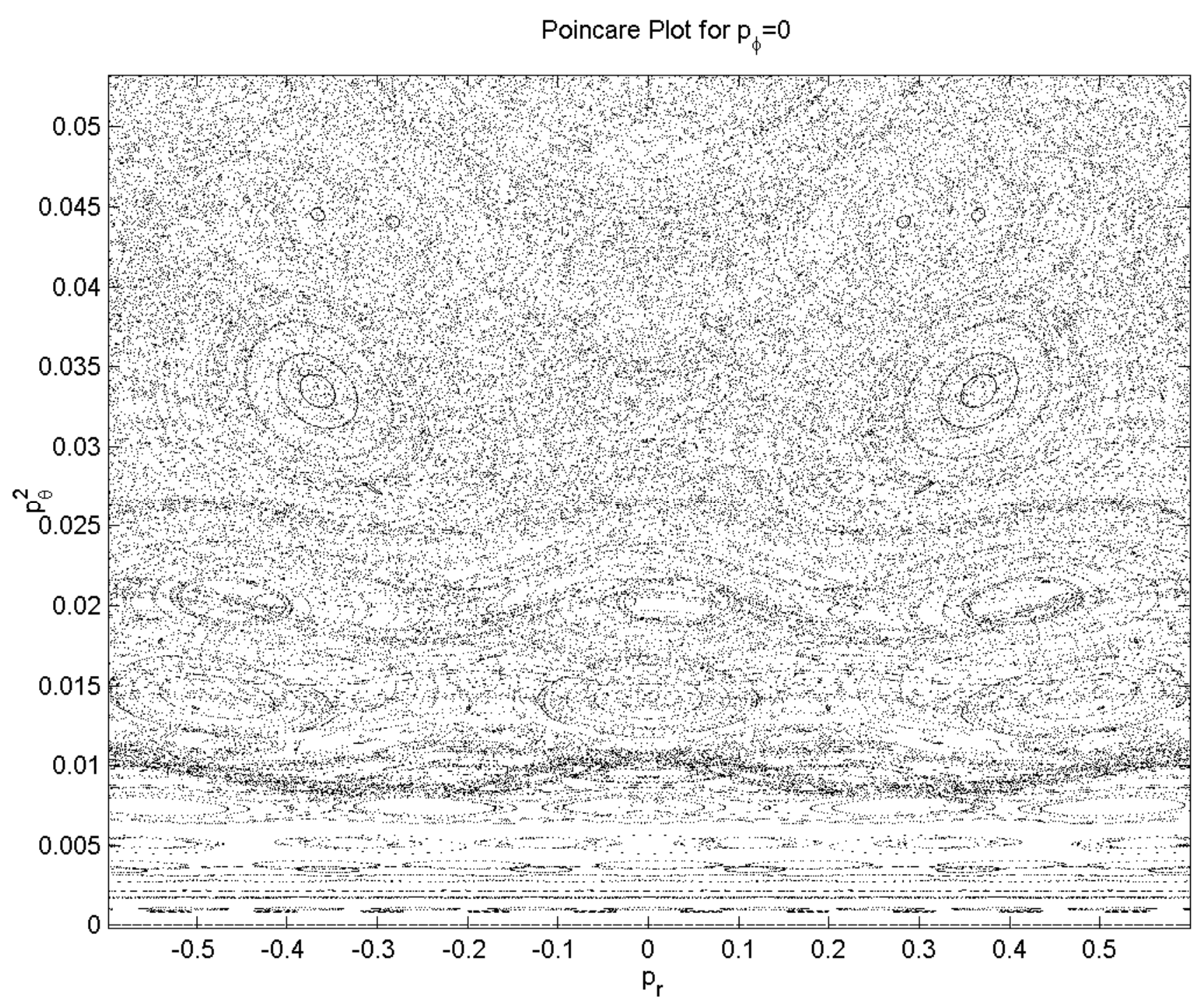}
\par\end{centering}

\caption{A zoomed-in image of the bottom portion of Figure \ref{fig:poinc_tot_bottom}.}

\label{fig:poinc_tot_bottom_zoomed_bot} 
\end{figure}

For the side slice, we could not use the condition of $p_{\theta}$
being \textit{exactly} zero. \ $p_{\phi}=0$ will be true for any
of the two-dimensional trajectories in planes passing through the
$\alpha$ axis. \
Since $p_{\alpha}$\ is calculated from $H=1$, most - if not all
- of the two-dimensional trajectories in our parameter scan will be
on planes passing through the $\alpha$\ axis (initial conditions
with $p_{\alpha}=0$\ are extremely unlikely). \ However, $p_{\theta}=0$\ would
only be true of a trajectory that remained in a cone rooted at the
origin. \ Hence we use the constraint of $p_{\theta}$\ being ``close''
to zero, instead of exactly zero to get an equivalent slice. We used
$p_{\theta}^{2}<0.0005$\ which produced about five-hundred thousand
points.

\begin{figure}[tbph]
\begin{centering}
\includegraphics[width=1\linewidth]{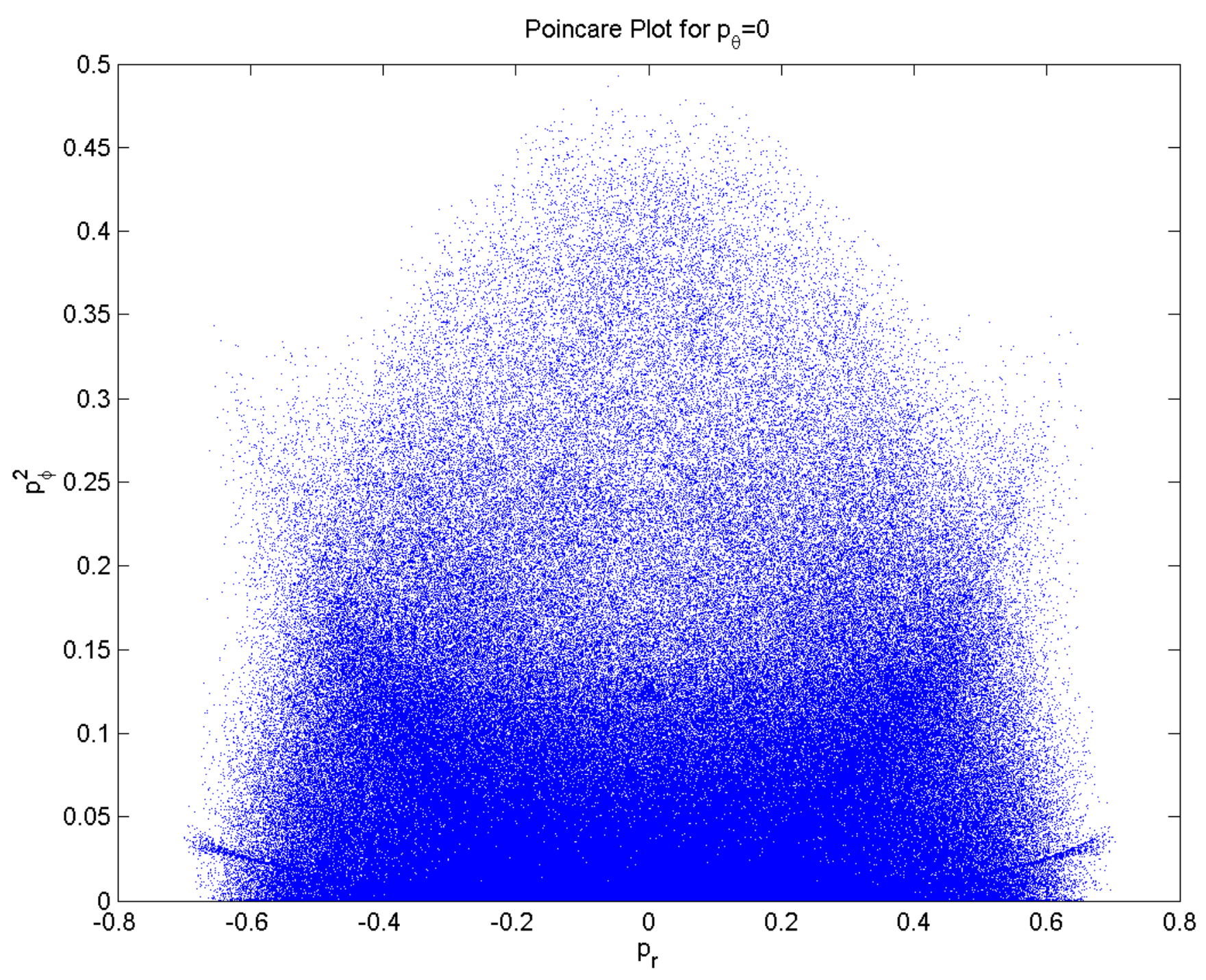}
\par\end{centering}

\caption{The side slice of the complete Poincare plot (\symbol{126}500,000
points). \ Note that this plot does not display the fractal-like
properties of the bottom slice (Figure \ref{fig:poinc_tot_bottom}).}

\label{fig:poinc_tot_side} 
\end{figure}

The first thing that we notice is that the side slice does not display
the same patterns and fractal-like properties as the bottom slice.
\ This could be due either to too few time steps or to a failure
to cover a sufficient range of initial conditions. \ It could also
possibly mean that spherical coordinates are not the best choice for
a generalization of Poincare plots; perhaps cylindrical coordinates
or another transformation would have produced more revealing results.
\ However it is also possible that the system simply does not demonstrate
any clearly definably pattern when sliced along the $p_{\theta}=0$\ plane.

We can get an idea of the true three-dimensional shape of this Poincare
graph by slicing the volume into multiple $p_{R}=k$\ planes, each
coloured according to density with brighter colours being higher densities.
\
Unfortunately this approach allows us to see little more than the
general ``shape'' of the plot. \ As the diagrammatic slices are
not enlightening we shall not present them here.

\bigskip{}

\section{Lyapunov Exponents}

\textcolor{black}{To compute Lyapunov exponents for the 4-body system,
\ we employ a method due to Benettin et.al.\cite{Benettin} for calculating
the largest Lyapunov exponent, using number of collisions instead
of time elapsed for calculating the largest Lyapunov exponents. First,
we take an initial deviation $\delta(0)$ away from a reference trajectory
that has already been calculated for $m$ collisions (the total number
of collisions). This perturbed trajectory is then calculated for $j$
collisions. Note that $m$ denotes the number of collision steps (with
$k$ denoting the step number) and $j$ is an integer that denotes
the sampling rate. For example, if we consider 100,000 collisions
sampled every 10th collision, then $m=100,000$, $j=10$, and $k$
would vary from 1 to 10,000. We assume that $m/j$ is an integer.}

\textcolor{black}{We then compare the reference and perturbed trajectory
after $j$ collisions (or after one step number) and calculate $\delta'(1)$,
the difference between our reference and perturbed trajectory. We
rescale this to have the same magnitude as $\delta(0)$, and then
repeat the process with $\delta(1)$ as the peturbation. At the $k$
th step we obtain the deviation $\delta'(k)$, and rescale it to have
the same magnitude as $\delta(0)$, which then becomes the new initial
deviation $\delta(k)$, calculated generally as shown in the expression
\[
\delta\left(k\right)=\frac{\delta'\left(k\right)}{d_{k}}\left\Vert \delta\left(0\right)\right\Vert \text{ \ \ \ \ where \ \ \ \ \ \ }d_{k}=\left\Vert \delta\left(k\right)\right\Vert 
\]
where $\delta'\left(k\right)$ is computed relative to the original
trajectory from $k=(n-1)$ to $k=n$ ($n=1,2,...,m/j$). This process
terminates for $k=m/j$.\ The l}argest Lyapunov exponent ($\lambda_{1}$
in our notation) is then obtained from 
\begin{equation}
\lambda_{1}=\lim_{n\rightarrow\infty}\left[\frac{1}{nj}\sum_{k=0}^{n}\ln\left(d_{k}\right)\right]\label{largelyap}
\end{equation}

Below is a table o\textcolor{black}{f calculated values for Lyapunov
exponents for figures shown throughout the paper. All of the values
below are considered accurate to $10^{-5}$, where all quantities
in the table should be multiplied by $10^{-4}$. The uncertainty arises
from the small fluctuations in the Lyapunov exponent graph which are
on the order of $10^{-5}$ for all graphs. Values for Lyapunov exponents
for bifurcated trajectories are not included because the graphs and
trajectories do not converge and do not exhibit Lyapunov-like behavior
and therefore the exponents are not relevant. Only 3 dimensional trajectories
are considered below so they are readily comparable to one another.
We have not computed Lyapunov exponents for the effective unequal-mass
2 dimensional (3-body) motions in figures from Figure \ref{fig:beg_2d}
(in which one coordinate and its conjugate momentum are fixed to always
be zero) since we are concerned here only with equal-mass 4-body case.
A study of the Lyapunov exponents for the unequal-mass 3-body case
remains an interesting subject for investigation.}

\smallskip{}

\begin{center}
\begin{tabular}{ll|ll|ll}
\multicolumn{6}{c}{Table of $\lambda_{1}$Values (after 200,000 collisions) $\times$
$10^{-4}$}\tabularnewline
\textcolor{black}{\footnotesize Fig. 4 ``Thick''Annulus} & \textcolor{black}{\footnotesize 121.4} & \textcolor{black}{\footnotesize Fig. $8{}^{\lyxmathsym{\dag}}$\ \ 1+1+1+1} & \textcolor{black}{\footnotesize 269.7} & \textcolor{black}{\footnotesize Fig. 13 1+1+2} & \textcolor{black}{\footnotesize 284.2}\tabularnewline
\textcolor{black}{\footnotesize Fig. 4 ``Thick'' Pretzel} & \textcolor{black}{\footnotesize 73.50} & \textcolor{black}{\footnotesize Fig. 9\ \  2+2} & \textcolor{black}{\footnotesize 130.7} & \textcolor{black}{\footnotesize Fig. 16 Poincare Pretzel} & \textcolor{black}{\footnotesize 4.700}\tabularnewline
\textcolor{black}{\footnotesize Fig. $5$ Chaotic Annulus} & \textcolor{black}{\footnotesize 213.8} & \textcolor{black}{\footnotesize Fig. 10\ \  3+1} & \textcolor{black}{\footnotesize 160.9} & \textcolor{black}{\footnotesize Fig. 17 Poincare Annulus} & \textcolor{black}{\footnotesize 26.02}\tabularnewline
\textcolor{black}{\footnotesize Fig. 6 Dense Pretzel} & \textcolor{black}{\footnotesize .2835} & \textcolor{black}{\footnotesize Fig. 11 \ \ 2+1+1} & \textcolor{black}{\footnotesize 233.6} & \textcolor{black}{\footnotesize Chaotic Trajectory{*}} & \textcolor{black}{\footnotesize 395.5}\tabularnewline
\textcolor{black}{\footnotesize Fig. 7 3D Periodic Trajectory} & \textcolor{black}{\footnotesize .3325} & \textcolor{black}{\footnotesize Fig. 12 \ \ 1+2+1} & \textcolor{black}{\footnotesize 113.3} &  & \tabularnewline
\end{tabular}
\par\end{center}

\smallskip{}

\dag{}The value of the Lyapunov exponent is for the perturbed plot
in the bottom left.

{*}initial conditions $\rho=0.1000$, $\beta=0.5000$, $\ \alpha=-0.2000$,
$\ p_{\rho}=0.2000$, $p_{\beta}=-0.2000$ and $\ p_{\alpha}=-0.3000$.

\noindent \bigskip{}

\bigskip{}

\ 
\begin{figure}[tbph]
\begin{centering}
\includegraphics[width=1\linewidth]{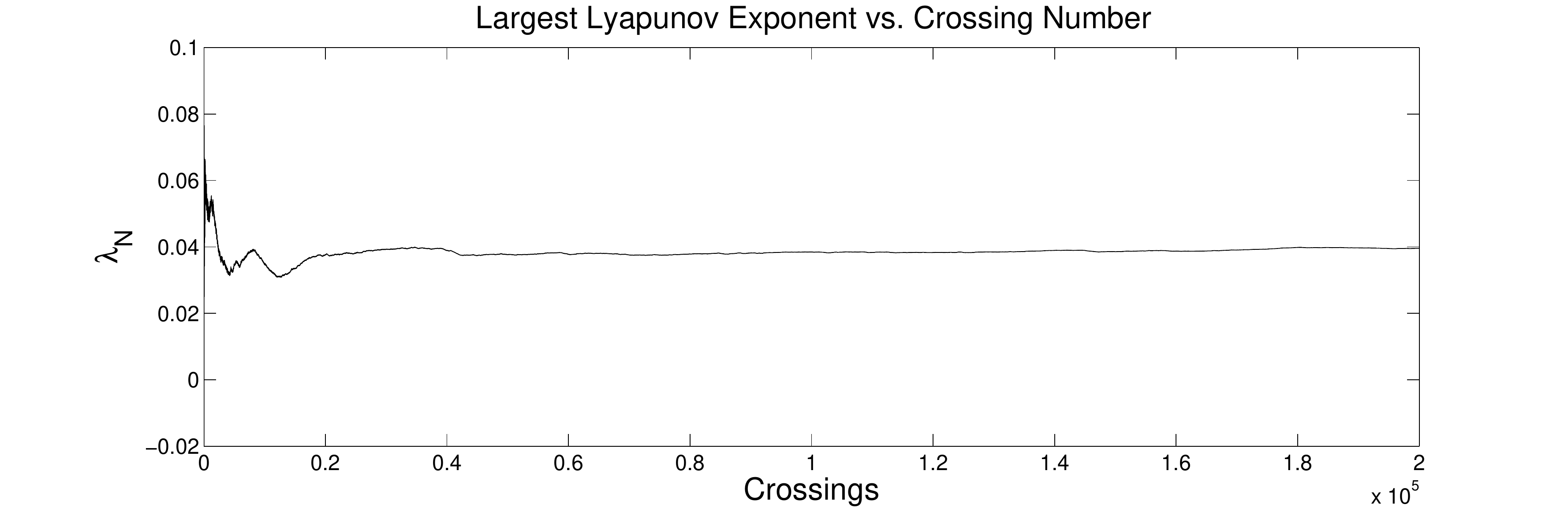}
\par\end{centering}

\caption{A plot of the largest Lyapunov exponent for the Chaotic Trajectory.
This trajectory was considered as a random set of initial conditions
that produced a chaotic trajectory in order to give a numerical reference
point against which we could compare other Lyapunov exponents.}

\label{ChaosLyap} 
\end{figure}

Looking at the Lyapunov exponents, a trend of values emerge. First,
trajectories that were considered to be mostly chaotic in nature (based
on the Lissajous figures accompanying them) have Lyapunov exponents
on the order of $10^{-2}.$ For trajectories that are periodic, the
Lyapunov exponents are on the order of $10^{-5}.$ For these small
Lyapunov exponents, the exact numerical values for the exponents are
more uncertain because the small fluctuations in these graphs are
on the order of the Lyapunov exponent. The mean value of the graphs
over the last hundred collisions are used to obtain a numerical estimate.
There also seems to be a range of intermediate values for Lyapunov
exponents for quasiperiodic trajectories in the range of $10^{-3}$to
$10^{-4}.$ 

One noteworthy feature of some Lyapunov graphs is orbital bifurcation,
in which the apparent stabilization of the largest Lyapunov exponent
is punctuated by a sudden increase toward a new value. We found this
to occur in a small number of graphs we analyzed\textcolor{black}{{}
(particularly figures 5 and 6 as shown). Further inspection of this
phenomenon indicated that the sudden increase was caused by differing
collision order between the unperturbed and perturbed trajectories,
causing the positions of the particles at a given collision number
to differ substantively. }

\textcolor{black}{The cause of this difference in collision order
was because the perturbed trajectory underwent an extra collision
that the unperturbed trajectory did not. Specifically in both cases,
the unperturbed trajectory approached a near 2-2 collision (where
a left and right pair of particles are about to collide). In the unperturbed
case, pair one crossed then pair two, whereas for the perturbed case
pair two crossed then pair one. Hence after the first collision the
unperturbed system is about to undergo a collision in pair two. Resetting
the perturbed trajectory after the first collision will then set it
on a course to have pair two collide, even though it has already undergone
a pair-two collision. Hence the perturbed trajectory undergoes pair-two
collision twice in a row without having had a pair-one collision,
putting it one collision behind the unperturbed trajectory. After
this the two systems undergo very similar motion, only with a relative
switch of two and one collision behind. This switch, along with the
lagging of the perturbed system, results in significant differences
between the two cases upon comparing the two systems. These differences
cause massive divergences in the Lyapunov exponents, seen as discontinuities
in the graphs.}

\textcolor{black}{Another unexpected phenomena within orbital bifurcation
is that in some cases, the bifurcated trajectories reconverged to
one another (as seen for figure 5). However, this is not always seen
on the time scale that was used for the analysis (as in figure 6).}

\textcolor{black}{In order to resolve this, the initial perturbation
size was decreased (from $10^{-6}$ to$10^{-10}$ in both cases) so
that this phenomenon disappeared and smooth Lyapunov graphs were obtained.
By reducing the perturbation size, the two trajectories stay close
enough to one another so the switch and extra collision do not occur.
However in both cases, each pair of particles are extremely close
to each other in the pair, suggesting that what should have occurred
is a simultaneous 2-2 crossing instead of the `C' motion of the particles
as is seen. Thus, this phenomena may simply be the fault of the numerics
and the sensitivity therein.}

\section{Conclusion}

We have carried out the first investigation of the non-relativistic
4-body problem for a one-dimensional self-gravitating system in the
equal mass case. \ This system is the largest value of $N$\ in
which the motion can be directly visualized in terms of the motion
of a single particle, referred to as the box particle, which is inside
a linear potential whose equipotential surfaces are a simplex that
has the shape of six square pyramids with their bottoms joined in
the shape of a cube. We showed how to classify the motions of this
system in terms of braid group operators, and were able to generalize
this classification to arbitrary values of $N$. \ 

We find that the trajectories of the box-particle form natural generalizations
of the 3-body case, thickening in one direction for small departures
from planar motion, in which the box-particle's position \textit{and}
momentum coordinates are initially zero. A common (and somewhat unexpected)
feature is that quasi-periodic motion can appear in certain two-dimensional
projections of the box-particle's trajectory whilst other projections
yield an apparently chaotic motion.

We generalized the Poincare plots of these trajectories to be three-dimensional,
where the radial momentum of the box-particle is plotted against the
two components of its angular momenta in spherical coordinates whenever
the box particle crosses a bisector of the equipotential simplex.
\
While plots for specific trajectories can be straightforwardly constructed,
a 3-dimensional Poincare plot over a large variety of initial conditions
proves extremely difficult to visualize in any practical sense. \ We
found that slices for constant azimuthal momenta yielded a fractal
pattern, but slices of constant polar momenta yielded no discernible
pattern.

\textcolor{black}{Finally, we computed the Lyapunov exponents using
the methods of ref. \cite{Goodings} for various trajectories and
computed the largest Lyapunov exponents. In general the Lyapunov exponent
graphs are quite stable and asymptote to values that seem to correspond
to their degree of stochasticity. These numeric results seem quite
reliable. Although, in future research analyzing Lyapunov exponents
using the collision method, it is important that perturbations are
kept sufficiently small to ensure no orbital bifurcation occurs.}

\smallskip{}

\noindent \begin{flushleft}
\textbf{\Large Acknowledgments}
\par\end{flushleft}{\Large \par}

\smallskip{}

This work was supported in part by the Natural Sciences and Engineering
Research Council of Canada. We are grateful to J. Emerson for helpful
discussions. \ We are grateful to Steve Blanchet for his contributions
to the early development of this work. We are also grateful to Marius
Oltean for his technical assistance while working on numerics.

\end{document}